\documentclass[aip,graphicx]{revtex4-1}
\usepackage[cp1251]{inputenc}
\usepackage[T2A]{fontenc}
\usepackage[english]{babel}
\usepackage{amssymb,latexsym,amsmath,amscd}
\usepackage{graphicx,color,framed}
\usepackage{xcolor}
\usepackage{siunitx} 
\begin{document}
\selectlanguage{english}
\title{Modified Poisson-Boltzmann equations and macroscopic forces in inhomogeneous ionic fluids}
\author{\firstname{Yury A.} \surname{Budkov}}
\email[]{ybudkov@hse.ru}
\affiliation{School of Applied Mathematics, HSE University, Tallinskaya st. 34, 123458 Moscow, Russia}
\affiliation{G.A. Krestov Institute of Solution Chemistry of the Russian Academy of Sciences, 153045, Akademicheskaya st. 1, Ivanovo, Russia}
\author{Andrei L. Kolesnikov}
\email[]{kolesnikov@inc.uni-leipzig.de}
\affiliation{Institut f\"ur Nichtklassische Chemie e.V., Permoserstr. 15, 04318 Leipzig, Germany}

\begin{abstract}
We propose a field-theoretical approach based on the thermodynamic perturbation theory and within it  derive a grand thermodynamic potential of the inhomogeneous ionic fluid as a functional of electrostatic potential for an arbitrary reference fluid system. We obtain a modified Poisson-Boltzmann equation as the Euler-Lagrange equation for the obtained functional. Applying Noether's theorem to this functional, we derive a general mean-field expression for the stress tensor consistent with the respective modified Poisson-Boltzmann equation. We derive a general expression for the macroscopic force acting on the dielectric or conductive body immersed in an ionic fluid. In particular, we derive a general mean-field expression for the disjoining pressure of an ionic fluid in a slit pore. We apply the developed formalism to describe three ionic fluid models of practical importance: nonpolarizable models (including the well-known Poisson-Boltzmann and Poisson-Fermi equations), polarizable models (ions carry nonzero permanent dipole or static polarizability), and models of ion-dipole mixtures (including the well-known Poisson-Boltzmann-Langevin equation). For these models, we obtain modified Poisson-Boltzmann equations and respective stress tensors, which could be valuable for different applications, where it is necessary to estimate the macroscopic forces acting on the dielectric or conductive bodies (electrodes, colloids, membranes, {\it etc}.) together with the local electrostatic potential (field) and ionic concentrations.
\end{abstract}
\maketitle
\section{Introduction}
Ionic fluids (plasma, electrolyte solutions, molten salts, and room temperature ionic liquids) have recently attracted a lot of researchers' and chemical engineers' attention. This was mostly caused by the use of ionic fluids in various applications, including lipid and ion-exchange membranes, biomacromolecules, colloids, batteries, fuel cells, supercapacitors, {\it etc}. In all these examples, ionic fluids, which either interact with the charged surface of the membrane, macromolecule, colloid, or electrode surface or are confined in charged nanopores, are strongly inhomogeneous. Due to the inhomogeneity of the ionic fluid, the local electrical neutrality is violated, so that its description requires numerical solution of one or another self-consistent field equation for the electrostatic potential with appropriate boundary conditions: the classical Poisson-Boltzmann (PB) equation or its generalizations. The latter are known in the literature as modified PB equations.

The modified PB equations obtained up to now take into account the individual properties of ionic species and the chemical specifics of the surface interacting with them. In this case it is necessary to especially emphasize the consideration of the non-zero size of ions (steric interactions)~\cite{kralj1996simple,borukhov1997steric,maggs2016general,kornyshev2007double}, their specific and structural interactions~\cite{goodwin2017mean,blossey2017structural}, polarizability/quadrupolarizability and permanent dipole moment of ionic species and solvent (co-solvent) molecules~\cite{frydel2011polarizable,budkov2020two,budkov2021theory,budkov2015modified,budkov2016theory,budkov2018theory,coalson1996statistical,abrashkin2007dipolar,iglivc2010excluded,buyukdagli2013microscopic,slavchov2014quadrupole}, surface charge regulaton~\cite{podgornik2018general}, specific ion-electrode interactions~\cite{budkov2018theory,uematsu2018effects} and electrostatic correlations~\cite{buyukdagli2020schwinger,de2020continuum,bazant2011double}. However, despite the variety of the modified PB equations, to the best of our knowledge, there is still no first-principles approaches allowing us to derive them from the unified standpoint and, simultaneously, take into account the aforementioned effects. Moreover, for applications, where it is required to calculate the macroscopic forces acting on conductive or dielectric bodies immersed in the ionic fluid (such as electrodes, colloid particles, membranes, micelles, dust particles, {\it etc}.), it is necessary to have a stress tensor consistent with the modified PB equation.

In this paper, we will present such a field-theoretical approach based on the thermodynamic perturbation theory allowing us to derive the grand thermodynamic potential of an inhomogeneous ionic fluid as a functional of the electrostatic potential for an arbitrary reference system, modified PB equation (as the Euler-Lagrange equation for this functional), and a stress tensor (with use of Noether's theorem). We will apply this approach to derivation of the modified PB equations and stress tensors for three types of models that can be employed in the applications: (1) unpolarizable ionic fluid models, (2) polarizable ionic fluid models, and (3) polarizable models of the ion-dipole mixture (solvent in salt models).

\section{Mean-field theory of ionic fluids}
\subsection{Model I. Nonpolarizable ionic fluid mean-field models}
Let us consider an ionic fluid -- a set of two types of ions with electric charges $q_{\pm}$ confined in a volume $V$ at a temperature $T$ contacting with an infinite reservoir with ions (bulk). We assume that the ions are immersed in a dielectric background of constant permittivity, $\varepsilon_b$. The ions interact with each other, in addition to the Coulomb interactions, via the short-range repulsive potentials $U_{\alpha\gamma}(\bold{r})$, 
where indexes $\alpha,\gamma=\pm$ enumerate the ions types. Thus, the grand partition function can be written as follows
\begin{equation}
\label{GPF}
\Xi=\sum\limits_{n_{+}=0}^{\infty}\sum\limits_{n_{-}=0}^{\infty}\frac{z_{+}^{n_{+}}z_{-}^{n_{-}}}{n_{+}!n_{-}!}Q_{n_{+}n_{-}},
\end{equation}
with the fugacities $z_{\pm}=\Lambda_{\pm}^{-3}e^{\beta\mu_{\pm}}\theta_{\pm}$ (expressed via the chemical potentials, $\mu_{\pm}$, the de Broglie thermal wavelengths, $\Lambda_\pm$, and internal partition functions, $\theta_{\pm}$, of the ions) of the ionic species and configuration integrals
\begin{equation}
Q_{n_{+}n_{-}}=\int d\Gamma_{n_+ n_-}\exp\left[-\beta U_{n_+ n_-}\right],
\end{equation}
where $\beta=(k_{B}T)^{-1}$ ($k_{B}$ is the Boltzmann constant, $T$ is the temperature) and
\begin{equation}
\int d\Gamma_{n_+ n_-}(\cdot)=\int..\int\prod\limits_{j=1}^{n_{+}}{d\bold{r}_{j}^{(+)}}\prod\limits_{k=1}^{n_{-}}{d\bold{r}_{k}^{(-)}}(\cdot).
\end{equation}
The total potential energy, $U_{n_+ n_-}$, of the interionic interactions can be written as a sum of several basic terms:
\begin{equation}
\label{hamilt}
U_{n_+ n_-}=U_{n_{+}n_{-}}^{(el)}+U_{n_+ n_-}^{(ev)}+U_{n_+ n_-}^{(ext)}-U_{self}.
\end{equation}
The first term in eq. (\ref{hamilt}) describes the total potential energy of electrostatic interactions
\begin{equation}
U_{n_+ n_-}^{(el)}=\frac{1}{2}\int d\bold{r}\int d\bold{r}^{\prime}\hat{\rho}(\bold{r})G(\bold{r}-\bold{r}^{\prime})\hat{\rho}(\bold{r}^{\prime})=\frac{1}{2}\left(\hat{\rho}G\hat{\rho}\right),
\end{equation}
where $G(\bold{r}-\bold{r}^{\prime})=1/4\pi \varepsilon_{b}|\bold{r}-\bold{r}^{\prime}|$ is the Green function of the Poisson equation; 
\begin{equation}
\hat{\rho}(\bold{r})=\hat{\rho}_{i}(\bold{r})+{\rho}_{ext}(\bold{r}),
\end{equation}
is the local charge density containing microscopic charge density of the ions
\begin{equation}
\hat{\rho}_{i}(\bold{r})=\sum\limits_{\alpha=\pm}q_{\alpha}\hat{c}_{\alpha}(\bold{r})
\end{equation}
with the microscopic ionic concentrations 
\begin{equation}
\hat{c}_{\alpha}(\bold{r})=\sum_{j=1}^{n_{\alpha}}\delta\left(\bold{r}-\bold{r}_{j}^{(\alpha)}\right)
\end{equation}
and the local charge density ${\rho}_{ext}(\bold{r})$ of the external charges. The second term in eq. (\ref{hamilt}) describes the total potential energy of the excluded volume interactions, i.e.
\begin{equation}
U_{n_+ n_-}^{(ev)}=\frac{1}{2}\sum\limits_{\alpha\gamma}^{}\left(\hat{c}_{\alpha}U_{\alpha\gamma}\hat{c}_{\gamma}\right)=\frac{1}{2}\int d\bold{r}\int d\bold{r}^{\prime}\sum\limits_{\alpha\gamma}^{}\hat{c}_{\alpha}(\bold{r})U_{\alpha\gamma}(\bold{r}-\bold{r}^{\prime})\hat{c}_{\gamma}(\bold{r}^{\prime}).
\end{equation}
The third term describes the interactions of ions with external fields with potential energies, $u_{\pm}(\bold{r})$: 
\begin{equation}
U_{n_+ n_-}^{(ext)}=\sum\limits_{\alpha}\int d\bold{r}\hat{c}_{\alpha}(\bold{r})u_{\alpha}(\bold{r})=\sum\limits_{\alpha}(\hat{c}_{\alpha}u_{\alpha}).
\end{equation}
The last term in eq. (\ref{hamilt}) is the self-energy of the ions determined by the following equation
$U_{self}=\left({n_{+}q_{+}^2}+{n_{-}q_{-}^2}\right)G(0)/2+({n_{+}U_{++}(0)}+{n_{-}U_{--}(0)})/2$.

Futher, using the standard Hubbard-Stratonovich transformations
\begin{equation}
\label{HS1}
e^{-\frac{\beta}{2}(\hat{\rho}G\hat{\rho})}=\int\frac{\mathcal{D}\varphi}{C}e^{-\frac{\beta}{2}(\varphi G^{-1}\varphi)+i\beta (\hat{\rho}\varphi)},    
\end{equation}
\begin{equation}
\label{HS2}
e^{-\frac{\beta}{2}\sum\limits_{\alpha\gamma}(\hat{c}_{\alpha}U_{\alpha\gamma}\hat{c}_{\gamma})}=\int \frac{\mathcal{D}\Phi}{\mathcal{N}_{U}}e^{-\frac{\beta}{2}\sum\limits_{\alpha\gamma}(\Phi_{\alpha}U^{-1}_{\alpha\gamma}\Phi_{\gamma})+i\beta \sum\limits_{\alpha}(\hat{c}_{\alpha}\Phi_{\alpha})},
\end{equation}
after some algebra, we obtain the following functional integral representation of the grand partition function
\begin{equation}
\Xi=\int \frac{\mathcal{D}\varphi}{C}e^{-\frac{\beta}{2}\left(\varphi G^{-1}\varphi\right)+i\beta(\rho_{ext}\varphi)}\Xi_{R}[\varphi],
\end{equation}
where $C=\int\mathcal{D}\varphi e^{-\frac{\beta}{2}\left(\varphi G^{-1}\varphi\right)}$ and $\mathcal{N}_{U}=\int{\mathcal{D}\Phi}e^{-\frac{\beta}{2}\sum\limits_{\alpha\gamma}(\Phi_{\alpha}U^{-1}_{\alpha\gamma}\Phi_{\gamma})}$ are the normalization constants of the Gaussian measures. The inverse electrostatic Green function has the following form
\begin{equation}
G^{-1}(\bold{r}-\bold{r}^{\prime})=-\varepsilon_{b}\nabla^2\delta(\bold{r}-\bold{r}^{\prime}).
\end{equation}
The grand partition function, $\Xi_{R}$, of the reference fluid with pure repulsive interactions between the particles can be written as the following functional integral
\begin{equation}
\Xi_{R}=\int \frac{\mathcal{D}\Phi}{\mathcal{N}_{U}}\exp\left[-\frac{\beta}{2}\sum\limits_{\alpha\gamma}(\Phi_{\alpha}U^{-1}_{\alpha\gamma}\Phi_{\gamma})+\sum\limits_{\alpha}\bar{z}_{\alpha}\int d\bold{r}e^{i\beta(q_{\alpha}\varphi+\Phi_{\alpha}+iu_{\alpha})}\right],
\end{equation}
where $\bar{z}_{\pm}=e^{\frac{\beta{U}_{\pm\pm}(0)}{2}+\frac{\beta q_{\pm}^2 G(0)}{2}}z_{\pm}\theta_{\pm}$ are the renormalized fugacities. The inverse operator, $U^{-1}$, is determined by the integral relation
\begin{equation}
\int d\bold{r}^{\prime\prime}\sum\limits_{\lambda} U_{\alpha\lambda}^{-1}(\bold{r}-\bold{r}^{\prime\prime})U_{\lambda\gamma}(\bold{r}^{\prime\prime}-\bold{r}^{\prime})=\delta_{\alpha\gamma}\delta(\bold{r}-\bold{r}^{\prime}).
\end{equation}

Also, using the cumulant expansion, we obtain
\begin{equation}
\Xi_{R}=\exp\left[\sum\limits_{n=1}^{\infty}\frac{\left<W^{n}[\Phi;\chi]\right>_{c}}{n!}\right],
\end{equation}
where we have introduced the functional 
\begin{equation}
\label{W}
W[\Phi;\chi]=\sum\limits_{\alpha}\bar{z}_{\alpha}\int d\bold{r} e^{i\beta \Phi_{\alpha}+i\beta\chi_{\alpha}}
\end{equation}
with the auxiliary variables $\chi_{\alpha}=q_{\alpha}\varphi+iu_{\alpha}$; and
the symbol $\left<(\cdot)\right>_{c}$ denotes the cumulant average \cite{kubo1962generalized}, so that the following relations take place
\begin{equation}
\left<W[\Phi;\chi]\right>_{c}=\left<W[\Phi;\chi]\right>,
\end{equation}
\begin{equation}
\left<W^{2}[\Phi;\chi]\right>_{c}=\left<W^{2}[\Phi;\chi]\right>-\left<W[\Phi;\chi]\right>^2,
\end{equation}
\begin{equation}
\left<W^{3}[\Phi;\chi]\right>_{c}=\left<W^{3}[\Phi;\chi]\right>-3\left<W^{2}[\Phi;\chi]\right>\left<W[\Phi;\chi]\right>+2\left<W[\Phi;\chi]\right>^3,...
\end{equation}
where we have introduced the notation for averaging over the Gaussian measure, i.e. 
\begin{equation}
\left<(\cdot)\right>= \int \frac{\mathcal{D}\Phi}{\mathcal{N}_{U}}\exp\left[-\frac{\beta}{2}\sum\limits_{\alpha\gamma}(\Phi_{\alpha}U^{-1}_{\alpha\gamma}\Phi_{\gamma})\right](\cdot).   
\end{equation}
Therefore, using the binomial formula, we obtain
\begin{equation}
\nonumber
\sum\limits_{n=1}^{\infty}\frac{\left<W^{n}[\Phi;\chi]\right>_{c}}{n!}=\sum\limits_{n=1}^{\infty}\sum\limits_{k=0}^{n}\frac{\bar{z}_{+}^{k}\bar{z}_{-}^{n-k}}{k!(n-k)!}\int d\bold{r}_{1}^{(+)}...\int d\bold{r}_{k}^{(+)}\int d\bold{r}_{1}^{(-)}...\int d\bold{r}_{n-k}^{(-)}
\end{equation}
\begin{equation}
\times \left<e^{i\beta \sum\limits_{s=1}^{k}\Phi_{+}\left(\bold{r}_{s}^{(+)}\right)+i\beta \sum\limits_{l=1}^{n-k}\Phi_{-}\left(\bold{r}_{s}^{(-)}\right)}\right>_{c} e^{i\beta \sum\limits_{s=1}^{k}\chi_{+}\left(\bold{r}_{s}^{(+)}\right)+i\beta \sum\limits_{l=1}^{n-k}\chi_{-}\left(\bold{r}_{l}^{(-)}\right)}.
\end{equation}
Then, after the calculations of the Gaussian functional integrals \cite{zinn2002quantum}, we get
\begin{equation}
\left<e^{i\beta \sum\limits_{s=1}^{k}\Phi_{+}(\bold{r}_{s}^{(+)})+i\beta \sum\limits_{l=1}^{n-k}\Phi_{-}(\bold{r}_{s}^{(-)})}\right>_{c}=e^{\frac{\beta k U_{++}(0)}{2}+\frac{\beta(n-k)U_{--}(0)}{2}}B^{nk}_{R}\left(\bold{r}_{1}^{(+)},...,\bold{r}_{k}^{(+)};\bold{r}_{1}^{(-)},...,\bold{r}_{n-k}^{(-)}\right),
\end{equation}
where we have introduced the cluster functions:
\begin{equation}
B_{R}^{11}\left(\bold{r}^{(+)}\right)=B_{R}^{10}\left(\bold{r}^{(-)}\right)=1,
\end{equation}
\begin{equation}
\nonumber
B_{R}^{22}\left(\bold{r}_{1}^{(+)},\bold{r}_{2}^{(+)}\right)=e^{-\beta U_{++}\left(\bold{r}_{1}^{(+)}-\bold{r}_{2}^{(+)}\right)}-1,~B_{R}^{21}\left(\bold{r}_{1}^{(+)};\bold{r}_{1}^{(-)}\right)=e^{-\beta U_{+-}\left(\bold{r}_{1}^{(+)}-\bold{r}_{1}^{(-)}\right)}-1,
\end{equation}
\begin{equation}
B_{R}^{20}\left(\bold{r}_{1}^{(-)},\bold{r}_{2}^{(-)}\right)=e^{-\beta U_{--}\left(\bold{r}_{1}^{(-)}-\bold{r}_{2}^{(-)}\right)}-1,
\end{equation}
\begin{equation}
\nonumber
B_{R}^{33}\left(\bold{r}_{1}^{(+)},\bold{r}_{2}^{(+)},\bold{r}_{3}^{(+)}\right)=e^{-\beta \left(U_{++}\left(\bold{r}_{1}^{(+)}-\bold{r}_{2}^{(+)}\right)+U_{++}\left(\bold{r}_{1}^{(+)}-\bold{r}_{3}^{(+)}\right)+U_{++}\left(\bold{r}_{2}^{(+)}-\bold{r}_{3}^{(+)}\right)\right)}
\end{equation}
\begin{equation}
\nonumber
-e^{-\beta U_{++}\left(\bold{r}_{1}^{(+)}-\bold{r}_{2}^{(+)}\right)}-e^{-\beta U_{++}\left(\bold{r}_{1}^{(+)}-\bold{r}_{3}^{(+)}\right)}-e^{-\beta U_{++}\left(\bold{r}_{2}^{(+)}-\bold{r}_{3}^{(+)}\right)}+2,
\end{equation}
\begin{equation}
\nonumber
B_{R}^{32}\left(\bold{r}_{1}^{(+)},\bold{r}_{2}^{(+)};\bold{r}_{1}^{(-)}\right)=e^{-\beta \left(U_{++}\left(\bold{r}_{1}^{(+)}-\bold{r}_{2}^{(+)}\right)+U_{+-}\left(\bold{r}_{1}^{(+)}-\bold{r}_{1}^{(-)}\right)+U_{+-}\left(\bold{r}_{2}^{(+)}-\bold{r}_{1}^{(-)}\right)\right)}
\end{equation}
\begin{equation}
\nonumber
-e^{-\beta U_{++}\left(\bold{r}_{1}^{(+)}-\bold{r}_{2}^{(+)}\right)}-e^{-\beta U_{+-}\left(\bold{r}_{1}^{(+)}-\bold{r}_{1}^{(-)}\right)}-e^{-\beta U_{+-}\left(\bold{r}_{2}^{(+)}-\bold{r}_{1}^{(-)}\right)}+2,
\end{equation}
\begin{equation}
\nonumber
B_{R}^{31}\left(\bold{r}_{1}^{(+)};\bold{r}_{1}^{(-)},\bold{r}_{2}^{(-)}\right)=e^{-\beta \left(U_{+-}\left(\bold{r}_{1}^{(+)}-\bold{r}_{1}^{(-)}\right)+U_{+-}\left(\bold{r}_{1}^{(+)}-\bold{r}_{2}^{(-)}\right)+U_{--}\left(\bold{r}_{1}^{(-)}-\bold{r}_{2}^{(-)}\right)\right)}
\end{equation}
\begin{equation}
\nonumber
-e^{-\beta U_{+-}\left(\bold{r}_{1}^{(+)}-\bold{r}_{1}^{(-)}\right)}-e^{-\beta U_{+-}\left(\bold{r}_{1}^{(+)}-\bold{r}_{2}^{(-)}\right)}-e^{-\beta U_{--}\left(\bold{r}_{1}^{(-)}-\bold{r}_{2}^{(-)}\right)}+2,
\end{equation}
\begin{equation}
\nonumber
B_{R}^{30}\left(\bold{r}_{1}^{(-)},\bold{r}_{2}^{(-)},\bold{r}_{3}^{(-)}\right)=e^{-\beta \left(U_{--}\left(\bold{r}_{1}^{(-)}-\bold{r}_{2}^{(-)}\right)+U_{--}\left(\bold{r}_{2}^{(-)}-\bold{r}_{3}^{(-)}\right)+U_{--}\left(\bold{r}_{1}^{(-)}-\bold{r}_{3}^{(-)}\right)\right)}
\end{equation}
\begin{equation}
-e^{-\beta U_{--}\left(\bold{r}_{1}^{(-)}-\bold{r}_{2}^{(-)}\right)}-e^{-\beta U_{--}\left(\bold{r}_{2}^{(-)}-\bold{r}_{3}^{(-)}\right)}-e^{-\beta U_{--}\left(\bold{r}_{1}^{(-)}-\bold{r}_{3}^{(-)}\right)}+2,...
\end{equation}
where index $k$ denotes the number of positively charged ions among the group of $n$ ions. At sufficiently small radii of the steric interactions, the cluster functions, $B^{nk}_{R}$, significantly deviate from zero, only when the positions of the particles are very close to each other. This allows us to adopt the following approximation for cumulant expansion
\begin{equation}
\nonumber
\sum\limits_{n=1}^{\infty}\frac{\left<W^{n}[\Phi;\chi]\right>_{c}}{n!}=\sum\limits_{n=1}^{\infty}\sum\limits_{k=0}^{n}\frac{\tilde{z}_{+}^{k}\tilde{z}_{-}^{n-k}}{k!(n-k)!}\int d\bold{r}_{1}^{(+)}..\int d\bold{r}_{k}^{(+)}\int d\bold{r}_{1}^{(-)}..\int d\bold{r}_{n-k}^{(-)} 
\end{equation}
\begin{equation}
\nonumber
\times B^{nk}_{R}\left(\bold{r}_{1}^{(+)},...,\bold{r}_{k}^{(+)};\bold{r}_{1}^{(-)},...,\bold{r}_{n-k}^{(-)}\right)e^{i\beta \sum\limits_{s=1}^{k}\chi_{+}\left(\bold{r}_{s}^{(+)}\right)+i\beta \sum\limits_{l=1}^{n-k}\chi_{-}\left(\bold{r}_{l}^{(-)}\right)}
\end{equation}
\begin{equation}
\nonumber
\approx \int d\bold{r}\sum\limits_{n=1}^{\infty}\sum\limits_{k=0}^{n}\frac{\tilde{z}_{+}^{k}\tilde{z}_{-}^{n-k}}{k!(n-k)!}e^{i\beta k\chi_{+}\left(\bold{r}\right)+i\beta(n-k) \chi_{-}\left(\bold{r}\right)} 
\end{equation}
\begin{equation}
\nonumber
\times \int d\bold{r}_{2}^{(+)}..\int d\bold{r}_{k}^{(+)}\int d\bold{r}_{1}^{(-)}..\int d\bold{r}_{n-k}^{(-)} B^{nk}_{R}\left(\bold{r}_{1}^{(+)},...,\bold{r}_{k}^{(+)};\bold{r}_{1}^{(-)},...,\bold{r}_{n-k}^{(-)}\right)
\end{equation}
\begin{equation}
\nonumber
=\int d\bold{r}\sum\limits_{n=1}^{\infty}\sum\limits_{k=0}^{n}b_{nk}{\tilde{z}_{+}^{k}\tilde{z}_{-}^{n-k}} e^{i\beta k\chi_{+}\left(\bold{r}\right)+i\beta(n-k) \chi_{-}\left(\bold{r}\right)},
\end{equation}
where $\tilde{z}_{\pm}=z_{\pm} e^{\beta q_{\pm}^2G(0)/2}\theta_{\pm}$ and 
\begin{equation}
b_{nk}=\frac{1}{k!(n-k)!}\int d\bold{r}_{2}^{(+)}...\int d\bold{r}_{k}^{(+)}\int d\bold{r}_{1}^{(-)}...\int d\bold{r}_{n-k}^{(-)} B^{nk}_{R}\left(\bold{r}_{1}^{(+)},...,\bold{r}_{k}^{(+)};\bold{r}_{1}^{(-)},...,\bold{r}_{n-k}^{(-)}\right)    
\end{equation}
are the cluster integrals for a two-component fluid \cite{hill1986introduction}. In other words, the cluster functions in this 'local' approximation can be approximated as
\begin{equation}
B^{nk}_{R}\left(\bold{r}_{1}^{(+)},...,\bold{r}_{k}^{(+)};\bold{r}_{1}^{(-)},...,\bold{r}_{n-k}^{(-)}\right)=k!(n-k)!b_{nk}\prod\limits_{j=2}^{k}\delta\left(\bold{r}_{j}^{(+)}-\bold{r}_{1}^{(+)}\right)\prod\limits_{l=1}^{n-k}\delta\left(\bold{r}_{l}^{(-)}-\bold{r}_{1}^{(+)}\right).
\end{equation}
Therefore, in the local approximation, the cumulant expansion is simply a cluster expansion \cite{hill1986introduction} of the bulk pressure as a function of shifted chemical potentials, $\bar{\mu}_{\pm}=\mu_{\pm}+i\chi_{\pm}+{q_{\pm}^2G(0)}/{2}+k_{B}T\ln\theta_{\pm}$, i.e.
\begin{equation}
\sum\limits_{n=1}^{\infty}\frac{\left<W^{n}[\Phi;\chi]\right>_{c}}{n!}\approx \beta\int d\bold{r}{P(T,\bar{\mu}_{+},\bar{\mu}_{-})}.
\end{equation}
Redefining the chemical potentials of the species $\bar{\mu}_{\pm}\rightarrow \bar{\mu}_{\pm}-{q_{\pm}^2G(0)}/{2}-k_{B}T\ln\theta_{\pm}$, we obtain the following local approximation for the grand partition function of the reference fluid
\begin{equation}
\label{loc_app}
\Xi_{R}=\exp\left[\beta\int d\bold{r} P(T,\bar{\mu}_{+},\bar{\mu}_{-})\right],    
\end{equation}
where $P\left(T,\bar{\mu}_{+},\bar{\mu}_{-}\right)$ is the reference fluid pressure as a function of the intrinsic chemical potentials 
$\bar{\mu}_{\pm}=\mu_{\pm}+iq_{\pm}\varphi-u_{\pm}$ of species and temperature. We would like to note that this approximation is in accordance with the thermodynamic relation $\Omega=-PV$ for the bulk systems.

Therefore, the grand partition function of the ionic fluid takes the form
\begin{equation}
\Xi=\int \frac{\mathcal{D}\varphi}{C}\exp\left[-\beta \Omega_{ef}[\varphi]\right],
\end{equation}
where we have introduced the following grand thermodynamic potential
\begin{equation}
\label{omega}
\Omega_{ef}[\varphi]=\frac{1}{2}\left(\varphi G^{-1}\varphi\right)-i(\rho_{ext}\varphi)-\int d\bold{r}P\left(T,\bar{\mu}_{+},\bar{\mu}_{-}\right),
\end{equation}
as a functional of the fluctuating electrostatic potential, $\varphi$. We would like to note that the functional (\ref{omega}) can be considered as the Landau-Ginzburg-Wilson effective Hamiltonian for the ionic fluid ($\varphi(\bold{r})$ is an order parameter) similar to those obtained within the similar approaches for simple fluids \cite{ivanchenko1984ginzburg,brilliantov1998effective,brilliantov2020molecular}.

Then, using the mean-field approximation, i.e. the functional derivative of the GTP to zero
\begin{equation}
\frac{\delta \Omega_{ef}}{\delta \varphi(\bold{r})}=0
\end{equation}
and introducing the self-consistent field potential (saddle-point) $\varphi^{(MF)}=i\psi$, we arrive at the following self-consistent field equation
\begin{equation}
\label{Poisson}
\Delta\psi(\bold{r})=-\frac{1}{\varepsilon_{b}}\left(\rho_{ext}(\bold{r})+q_{+}\bar{c}_{+}(\bold{r})+q_{-}\bar{c}_{-}(\bold{r})\right)
\end{equation}
where the local concentrations, $\bar{c}_{\pm}(\bold{r})$, are determined by the following thermodynamic relations
\begin{equation}
\bar{c}_{\gamma}(\bold{r})=\frac{\delta\Omega_{ef}}{\delta{u}_{\gamma}(\bold{r})}=\frac{\partial{P\left(T,\bar{\mu}_{+}(\bold{r}),\bar{\mu}_{-}(\bold{r})\right)}}{\partial{\bar{\mu}_{\gamma}}}=c_{\gamma}\left(T,\bar{\mu}_{+}(\bold{r}),\bar{\mu}_{-}(\bold{r})\right),
\end{equation} 
where $c_{\gamma}\left(T,\bar{\mu}_{+}(\bold{r}),\bar{\mu}_{-}(\bold{r})\right)$ are the concentrations of the reference fluid species. Thus, within the mean-field approximation, the ionic concentrations are determined as the molecular concentrations of the reference fluid and depend on the position, $\bold{r}$, via the intrinsic chemical potentials $\bar{\mu}_{\pm}(\bold{r})=\mu_{\pm}-q_{\pm}\psi(\bold{r})-u_{\pm}(\bold{r})$. Solving eq. (\ref{Poisson}) with the appropriate boundary conditions, we can calculate the GTP in the mean-field approximation as
\begin{equation}
\label{Omega33}
\Omega=\Omega_{ef}[\varphi^{(MF)}]=\int d\bold{r}\left(-\frac{\varepsilon_{b}(\nabla{\psi})^2}{2}+\rho_{ext}\psi-P(T,\bar{\mu}_{+},\bar{\mu}_{-})\right).
\end{equation}
However, when the ionic fluid adjoins with conductors possessing fixed electric potentials or surface charges and there are no fixed volume charges, we can use the following functional 
\begin{equation}
\Omega[\psi]=\int d\bold{r}\left(-\frac{\varepsilon_{b}(\nabla{\psi})^2}{2}-P(T,\bar{\mu}_{+},\bar{\mu}_{-})\right),
\end{equation}
related to the following self-consistent field equation 
\begin{equation}
\label{Poisson_eq_3}
\Delta\psi(\bold{r})=-\frac{1}{\varepsilon_{b}}\left(q_{+}\bar{c}_{+}(\bold{r})+q_{-}\bar{c}_{-}(\bold{r})\right).
\end{equation}

Note that equation (\ref{Poisson}) can be obtained directly by varying functional (\ref{Omega33}) with respect to the potential, $\psi$. In other words, functional (\ref{Omega33}) reaches the extreme value at solutions to eq. (\ref{Poisson}). This idea was used in papers \cite{budkov2018theory,budkov2016theory,maggs2016general,budkov2020two,blossey2017structural} to obtain the self-consistent field equations by varying the GTP rewritten as a functional of electrostatic potential \cite{budkov2016theory,maggs2016general} and potentials of specific interactions \cite{budkov2018theory,blossey2017structural} by using the local Legendre transformation. In this paper, we justify this heuristic approach, starting from the grand partition function of ionic fluid.

In order to apply this theory to inhomogeneous electrolyte solutions or room temperature ionic liquids, it is necessary to specify the dependence of osmotic pressure on chemical potentials, $P(T,\bar{\mu}_{+},\bar{\mu}_{-})$. For the case of an ideal gas reference system, for which
\begin{equation}
P(T,\bar{\mu}_{+},\bar{\mu}_{-})=\frac{k_{B}T}{\Lambda_{+}^3}e^{\beta\bar{\mu}_{+}}+\frac{k_{B}T}{\Lambda_{-}^3}e^{\beta\bar{\mu}_{-}},
\end{equation}
where $\Lambda_{\pm}$ are the thermal de Broglie wavelengths, we obtain the classical Poisson-Boltzmann equation
\begin{equation}
\label{Poisson-Boltzmann}
\Delta\psi(\bold{r})=-\frac{1}{\varepsilon_{b}}\left(q_{+}c_{+,b}e^{-\beta q_{+}\psi(\bold{r})}+q_{-}c_{-,b}e^{-\beta q_{-}\psi(\bold{r})}\right),
\end{equation}
where we used the relations for bulk chemical potentials $\mu_{\pm}=k_{B}T\ln\left(\Lambda_{\pm}^3c_{\pm,b}\right)$, with $c_{\pm,b}$ being the ionic concentrations in the bulk (where $\psi=0$). Another widely used reference fluid system is a symmetric lattice gas, for which such dependence has the form \cite{hill1986introduction,maggs2016general}
\begin{equation}
P(T,\bar{\mu}_{+},\bar{\mu}_{-})=\frac{k_{B}T}{v}\ln\left(1+e^{\beta\bar{\mu}_{+}}+e^{\beta\bar{\mu}_{-}}\right),
\end{equation}
where $v$ is the lattice cell volume. In this case, when $q_{\pm}=\pm q$ and $\mu_{\pm}=k_{B}T\ln{cv/(1-2cv)}$ the self-consistent field equation transforms into the so-called Poisson-Fermi equation~\cite{kornyshev2007double,bazant2009towards,borukhov1997steric}
\begin{equation}
\Delta\psi(\bold{r})=\frac{2c q}{\varepsilon_{b}}\frac{\sinh{\beta q\psi(\bold{r})}}{1+2cv\left(\cosh\beta q\psi(\bold{r})-1\right)},
\end{equation}  
where $c$ is the bulk ionic concentration. In order to take into account the ionic size asymmetry, we can use the asymmetric lattice gas model as the reference system, which was systematically considered in paper~\cite{maggs2016general} in the context of the electric double layer theory. To take into account the ionic size asymmetry, we can also use the hard sphere mixture, for which the local pressure and chemical potentials can be obtained from the following excess free energy density within the Percus-Yevick approximation:
\begin{equation}
\label{PY}
\beta f_{ex} = -n_0 \ln(1-n_3) + \frac{n_1 n_2}{1 - n_3} + \frac{n_2^3}{24\pi(1-n_3)^2}
\end{equation}
or using the Carnahan-Starling approximation~\cite{roth2010fundamental,maggs2016general}. The auxiliary variables are: $n_0 = \bar{c}_{+} + \bar{c}_{-}$, $n_1 = 1/2 ~ d_+ \bar{c}_{+} + 1/2 ~ d_- \bar{c}_{-}$, $n_2 = \pi~d_+^2 \bar{c}_{+} +\pi ~ d_-^2 \bar{c}_{-}$, $n_3 = \pi/6 ~ d_+^3 \bar{c}_{+} + \pi/6 ~ d_-^3 \bar{c}_{-}$. The pressure and chemical potentials of species ($\alpha = \pm$) can be derived via standard thermodynamic relations~\cite{roth2002fundamental}:
\begin{equation}
   \label{press}
    \beta P = n_0 \frac{1 + n_3 + n_3^2}{(1-n_3)^3} + \frac{1/12\pi~n_2^3 + n_1 n_2 (1 - n_3) -3 n_0 n_3}{(1-n_3)^3},
\end{equation}
\begin{equation}
\label{mu}
\beta\bar{\mu}_{\alpha} =\ln\frac{\bar{c}_{\alpha}\Lambda_{\alpha}^3}{1-n_3}+\frac{n_2 d_{\alpha}}{2(1-n_3)}+\pi d_{\alpha}^2\left(\frac{n_1}{1-n_3} + \frac{n_2^2}{8\pi (1-n_3)^2}\right)+\frac{\beta \pi d_{\alpha}^3P}{6},
\end{equation}
where $d_{\alpha}$ are the ionic effective diameters. 

\subsection{Model II. Polarizable ionic fluid mean-field models}
Now let us consider an ionic fluid, whose ions carry charges $q_{\pm}+e_{\pm}$ at the center and additional 'peripheral' charges $-e_{\pm}$ at a fluctuating distance from it, so that the total charges of the ions are $q_{\pm}$. The fluctuating distance, $\bold{\xi}$, is described by the distribution functions $g_{\pm}(\bold{\xi})$. This model can be used to describe room temperature ionic liquids and molten salts with polarizable/polar ions. A similar 'nonlocal' description of the bound charge of ions and molecules was also presented in papers~\cite{buyukdagli2013microscopic,budkov2019statistical,budkov2020statistical,budkov2018nonlocal}. The configuration integrals in the grand partition function (\ref{GPF}) take the form
\begin{equation}
Q_{n_+ n_-}=\int d\Gamma_{n_+ n_-}\exp\left[-\beta U_{n_+ n_-}\right],
\end{equation}
where
\begin{equation}
\int d\Gamma_{n_+ n_-}(\cdot)=\int..\int\prod\limits_{j=1}^{n_{+}}{d\bold{r}_{j}^{(+)}}d\bold{\xi}^{(+)}_{j}g_{+}\left(\bold{\xi}^{(+)}_{j}\right)\prod\limits_{k=1}^{n_{-}}{d\bold{r}_{k}^{(-)}}d\bold{\xi}^{(-)}_{k}g_{-}\left(\bold{\xi}^{(-)}_{k}\right)(\cdot).
\end{equation}
The total potential energy is determined by the same expression (\ref{hamilt}) with the electrostatic energy expressed via the following local charge density
\begin{equation}
\hat{\rho}(\bold{r})=\hat{\rho}_{i}(\bold{r})+{\rho}_{ext}(\bold{r}),
\end{equation}
where
\begin{equation}
\label{rho_i}
\hat{\rho}_{i}(\bold{r})=\sum\limits_{\alpha}q_{\alpha}\hat{c}_{\alpha}(\bold{r})+\sum\limits_{\alpha=\pm}\sum\limits_{j_{\alpha}=1}^{n_{\alpha}}e_{\alpha}\left(\delta\left(\bold{r}-\bold{r}^{(\alpha)}_{j_{\alpha}}\right)-\delta\left(\bold{r}-\bold{r}^{(\alpha)}_{j_{\alpha}}-\bold{\xi}_{j_{\alpha}}^{(\alpha)}\right)\right)
\end{equation}
is the microscopic charge density of the ions taking into account their polarizability (the second term).

Performing the same transformations as in Subsection A, we arrive at
\begin{equation}
\Xi=\int \frac{\mathcal{D}\varphi}{C}e^{-\frac{\beta}{2}\left(\varphi G^{-1}\varphi\right)+i\beta(\rho_{ext}\varphi)}\Xi_{R},
\end{equation}
where we have introduced the reference grand partition function
\begin{equation}
\Xi_{R}=\int \frac{\mathcal{D}\Phi}{\mathcal{N}_{U}}\exp\left[-\frac{\beta}{2}\sum\limits_{\alpha\gamma}(\Phi_{\alpha}U^{-1}_{\alpha\gamma}\Phi_{\gamma})+W[\Phi;\chi]\right]
\end{equation}
with the functional $W[\Phi;\chi]$ determined by eq. (\ref{W}) with
\begin{equation}
\chi_{\alpha}(\bold{r})=q_{\alpha}\varphi(\bold{r})+iu_{\alpha}(\bold{r})+k_{B}T\ln\left[\int d\bold{\xi}g_{\alpha}(\bold{\xi})e^{i\beta e_{\alpha}\left(\varphi(\bold{r})-\varphi(\bold{r}+\bold{\xi})\right)}\right].
\end{equation}
Then, performing the same calculations based on the cumulant expansion and adopting the same approximations as in Subsection A, we obtain the GTP in the absence of external volume electric charges
\begin{equation}
\label{Omega4}
\Omega[\psi]=\int d\bold{r}\left(-\frac{\varepsilon_{b}(\nabla{\psi})^2}{2}-P(T,\bar{\mu}_{+},\bar{\mu}_{-})\right),
\end{equation}
where
\begin{equation}
\bar{\mu}_{\alpha}(\bold{r})=\mu_{\alpha}-q_{\alpha}\psi(\bold{r})+k_{B}T\ln\left[\int d\bold{\xi}g_{\alpha}(\bold{\xi})e^{-\beta e_{\alpha}\left(\psi(\bold{r})-\psi(\bold{r}+\bold{\xi})\right)}\right].
\end{equation}
Now we would like to specify the probability distribution functions $g_{\alpha}(\bold{\xi})$. We consider the Gaussian distributions
\begin{equation}
\label{PDE}
g_{\alpha}(\bold{\xi})=(2\pi\sigma_{\alpha}^2)^{-3/2}\int\frac{d\bold{n}_{\alpha}}{4\pi}\exp\left[-\frac{(\bold{\xi}-\bold{s}_{\alpha})^2}{2\sigma_{\alpha}^2}\right],
\end{equation}
averaged over orientations of unit vectors $\bold{n}_{\alpha}=\bold{s}_{\alpha}/|\bold{s}_{\alpha}|$, where $\bold{s}_{\alpha}$ is the vector of the average distance between charges $\pm e_{\alpha}$. In this case, we obtain
\begin{equation}
\nonumber
\int d\bold{\xi}g_{\alpha}(\bold{\xi})e^{-\beta e_{\alpha}\left(\psi(\bold{r})-\psi(\bold{r}+\bold{\xi})\right)}=\int\frac{d\bold{n}_{\alpha}}{4\pi}\int\frac{d\bold{\xi}}{(2\pi\sigma_{\alpha}^2)^{3/2}}e^{-\frac{(\bold{\xi}-\bold{s}_{\alpha})^2}{2\sigma_{\alpha}^2}}e^{-\beta e_{\alpha}\left(\psi(\bold{r})-\psi(\bold{r}+\bold{\xi})\right)}
\end{equation}
\begin{equation}
\approx \int\frac{d\bold{n}_{\alpha}}{4\pi}\int\frac{d\bold{\xi}}{(2\pi\sigma_{\alpha}^2)^{3/2}}e^{-\frac{(\bold{\xi}-\bold{s}_{\alpha})^2}{2\sigma_{\alpha}^2}}e^{-\beta e_{\alpha}\bold{\xi}\nabla\psi(\bold{r})}= \frac{\sinh(\beta p_{\alpha}|\nabla \psi(\bold{r})|)}{\beta p_{\alpha}|\nabla \psi(\bold{r})|}e^{\frac{\gamma_{\alpha}(\nabla\psi(\bold{r}))^2}{2k_{B}T}},
\end{equation}
where we have used the approximation $\psi(\bold{r}+\bold{\xi})\approx \psi(\bold{r})+\bold{\xi}\nabla\psi(\bold{r})$ and introduced the permanent dipole moments, $p_{\alpha}=e_{\alpha}|\bold{s}_{\alpha}|$, and static polarizabilities, $\gamma_{\alpha}=\beta\sigma_{\alpha}^2e_{\alpha}^2$, of the ions. Therefore, we obtain the following expressions for the intrinsic chemical potentials of the species
\begin{equation}
\label{chem_pot}
\bar{\mu}_{\alpha}(\bold{r})=\mu_{\alpha}-q_{\alpha}\psi(\bold{r})+\frac{\gamma_{\alpha}(\nabla\psi(\bold{r}))^2}{2}+k_{B}T\ln\left[\frac{\sinh(\beta p_{\alpha}|\nabla \psi(\bold{r})|)}{\beta p_{\alpha}|\nabla \psi(\bold{r})|}\right].
\end{equation}
Variation of the functional (\ref{Omega4}) with respect to the electrostatic potential, $\psi(\bold{r})$, with an account for eq. (\ref{chem_pot}) leads to the following self-consistent field equation
\begin{equation}
\label{scf_eq2}
-\nabla\cdot(\epsilon(\bold{r})\nabla\psi(\bold{r}))=q_{+}\bar{c}_{+}(\bold{r})+q_{-}\bar{c}_{-}(\bold{r}),
\end{equation}
where
\begin{equation}
\label{eps}
\epsilon(\bold{r})=\varepsilon_{b} + \sum\limits_{\alpha=\pm}\left(\gamma_{\alpha}+\frac{p_{\alpha}^2}{k_{B}T}\frac{L(\beta p_{\alpha}\mathcal{E}(\bold{r}))}{\beta p_{\alpha}\mathcal{E}(\bold{r})}\right)\bar{c}_{\alpha}(\bold{r})  
\end{equation}
is the effective permittivity of the medium, $\mathcal{E}(\bold{r})=|\nabla \psi(\bold{r})|$ is the electric field absolute value, $L(x)=\coth x-1/x$ is the Langevin function. We would like to mention that eqs. (\ref{scf_eq2}-\ref{eps}), in case of the symmetric lattice gas model, were used in ref. \cite{budkov2021theory} to model the ionic liquids with a polarizable/polar cation on a charged electrode. A similar theory was also used to model two-component electrolyte solution with strongly polar cations on charged electrode \cite{budkov2020two}.

\subsection{Model III. Mean-field ion-dipole mixture models (solvent in salt models)}
Now let us consider the case of an ionic fluid with an admixture of electrically neutral but polarizable molecules. This model can be used to describe room temperature ionic fluids and molten salts with an additive of polar solvents (solvent in salt) or electrolyte solutions with an admixture of highly polar molecules, such as amino acids. As above, we assume that the ions carry charges $q_{\pm}+e_{\pm}$ at the center and additional charges $-e_{\pm}$ at a fluctuating distance from it. The neutral molecules also have charge $e_{0}$ at the center and charge $-e_{0}$ placed at a fluctuating distance from the center. The probability distribution functions $g_{\alpha}(\bold{\xi})$ ($\alpha=\pm,0$) are the same averaged Gaussian functions (\ref{PDE}). The grand partition function, thus, can be written as follows
\begin{equation}
\label{GPF}
\Xi=\sum\limits_{n_{+}=0}^{\infty}\sum\limits_{n_{-}=0}^{\infty}\sum\limits_{n_{0}=0}^{\infty}\frac{z_{+}^{n_{+}}z_{-}^{n_{-}}z_0^{n_0}}{n_{+}!n_{-}!n_{0}!}Q_{n_{+}n_{-}n_0}
\end{equation}
with the fugacities, $z_{\alpha}=\Lambda_{\alpha}^{-3}e^{\beta\mu_{\alpha}}\theta_{\alpha}$, ($\alpha=\pm,0$) of the ionic species and the configuration integrals
\begin{equation}
Q_{n_{+}n_{-}n_0}=\int d\Gamma_{n_{+}n_{-}n_0} \exp\left[-\beta U_{n_{+}n_{-}n_0}\right],
\end{equation}
\begin{multline}
\int d\Gamma_{n_{+}n_{-}n_0}(\cdot)=\int..\int\prod\limits_{j=1}^{n_{+}}{d\bold{r}_{j}^{(+)}}d\bold{\xi}^{(+)}_{j}g_{+}\left(\bold{\xi}^{(+)}_{j}\right)\\\times\prod\limits_{k=1}^{n_{-}}{d\bold{r}_{k}^{(-)}}d\bold{\xi}^{(-)}_{k}g_{-}\left(\bold{\xi}^{(-)}_{k}\right)\prod\limits_{l=1}^{n_{0}}{d\bold{r}_{l}^{(0)}}d\bold{\xi}^{(0)}_{l}g_{0}\left(\bold{\xi}^{(0)}_{l}\right)(\cdot),
\end{multline}
where the total potential energy is determined by the same eq. (\ref{hamilt}). The total local charge density, in this case, can be written in the form
\begin{equation}
\hat{\rho}(\bold{r})=\hat{\rho}_{i}(\bold{r})+\hat{\rho}_{d}(\bold{r})+{\rho}_{ext}(\bold{r}),
\end{equation}
where $\hat{\rho}_{i}(\bold{r})$ is already defined above by eq. (\ref{rho_i}) and 
\begin{equation}
\hat{\rho}_{d}(\bold{r})=\sum\limits_{j=1}^{N_{0}}e_{0}\left(\delta\left(\bold{r}-\bold{r}^{(0)}_j\right)-\delta\left(\bold{r}-\bold{r}^{(0)}_{j}-\bold{\xi}_{j}^{(0)}\right)\right)
\end{equation}
is the microscopic density of the bound charge of the molecules. Then, performing the same transformations as in Subsection A, we obtain
\begin{equation}
\Xi=\int \frac{\mathcal{D}\varphi}{C}e^{-\frac{\beta}{2}\left(\varphi G^{-1}\varphi\right)+i\beta(\rho_{ext}\varphi)}\Xi_{R}[\varphi],
\end{equation}
where we have introduced the reference grand partition function
\begin{equation}
\Xi_{R}=\int \frac{\mathcal{D}\Phi}{\mathcal{N}_{U}}\exp\left[-\frac{\beta}{2}\sum\limits_{\alpha\gamma}(\Phi_{\alpha}U^{-1}_{\alpha\gamma}\Phi_{\gamma})+W[\Phi;\chi]\right]
\end{equation}
with the functional
\begin{equation}
\label{W2}
W[\Phi;\chi]=\sum\limits_{\alpha=\pm,0}\bar{z}_{\alpha}\int d\bold{r} e^{i\beta \Phi_{\alpha}+i\beta\chi_{\alpha}},
\end{equation}
where
\begin{equation}
\chi_{\alpha}(\bold{r})=q_{\alpha}\varphi(\bold{r})+iu_{\alpha}(\bold{r})+k_{B}T\ln\left[\int d\bold{\xi}g_{\alpha}(\bold{\xi})e^{i\beta e_{\alpha}\left(\varphi(\bold{r})-\varphi(\bold{r}+\bold{\xi})\right)}\right],~\alpha=\pm
\end{equation}
\begin{equation}
\chi_{0}(\bold{r})=iu_{0}(\bold{r})+k_{B}T\ln\left[\int d\bold{\xi}g_{0}(\bold{\xi})e^{i\beta e_{0}\left(\varphi(\bold{r})-\varphi(\bold{r}+\bold{\xi})\right)}\right].
\end{equation}
After the same approximation, as in Subsection B, in the absence of external volume charge, we arrive at the following GTP
\begin{equation}
\label{Omega5}
\Omega[\psi]=\int d\bold{r}\left(-\frac{\varepsilon_{b}(\nabla{\psi})^2}{2}-P(T,\bar{\mu}_{+},\bar{\mu}_{-},\bar{\mu}_{0})\right)
\end{equation}
and self-consistent field equation
\begin{equation}
-\nabla\cdot(\epsilon(\bold{r})\nabla\psi(\bold{r}))=q_{+}\bar{c}_{+}(\bold{r})+q_{-}\bar{c}_{-}(\bold{r}),
\end{equation}
where
\begin{equation}
\epsilon(\bold{r})=\varepsilon_{b} + \sum\limits_{\alpha=\pm,0}\left(\gamma_{\alpha}+\frac{p_{\alpha}^2}{k_{B}T}\frac{L(\beta p_{\alpha}\mathcal{E}(\bold{r}))}{\beta p_{\alpha}\mathcal{E}(\bold{r})}\right)\bar{c}_{\alpha}(\bold{r})  
\end{equation}
is the effective permittivity taking into account polarizability of the ions and molecules and
\begin{equation}
\label{chem_pot2}
\bar{\mu}_{\alpha}(\bold{r})=\mu_{\alpha}-q_{\alpha}\psi(\bold{r})+\frac{\gamma_{\alpha}(\nabla\psi(\bold{r}))^2}{2}+k_{B}T\ln\left[\frac{\sinh(\beta p_{\alpha}|\nabla \psi(\bold{r})|)}{\beta p_{\alpha}|\nabla \psi(\bold{r})|}\right],~\alpha=\pm,
\end{equation}
\begin{equation}
\label{chem_pot3}
\bar{\mu}_{0}(\bold{r})=\mu_{0}+\frac{\gamma_{0}(\nabla\psi(\bold{r}))^2}{2}+k_{B}T\ln\left[\frac{\sinh(\beta p_{0}|\nabla \psi(\bold{r})|)}{\beta p_{0}|\nabla \psi(\bold{r})|}\right]
\end{equation}
are the intrinsic chemical potentials, where $p_0=e_0|\bold{s}_0|$ and $\gamma_0=\beta\sigma_0^2e_0^2$ are the permanent dipole moment and static polarizability of the molecules, respectively. The local concentrations of the species are
\begin{equation}
\bar{c}_{\alpha}(\bold{r})=\frac{\partial{P}}{\partial{\bar{\mu}_{\alpha}(\bold{r})}}=c_{\alpha}\left(T,\bar{\mu}_{+}(\bold{r}),\bar{\mu}_{-}(\bold{r}),\bar{\mu}_{0}(\bold{r})\right),~\alpha=\pm,0.
\end{equation} 

We would like to note that in case when $\gamma_{\pm}=\gamma_{0}=p_{\pm}=0$ and $p_{0}=p\neq 0$ and
\begin{equation}
P(T,\bar{\mu}_{+},\bar{\mu}_{-},\bar{\mu}_{0})=\frac{k_{B}T}{\Lambda_{+}^3}e^{\beta\bar{\mu}_{+}}+\frac{k_{B}T}{\Lambda_{-}^3}e^{\beta\bar{\mu}_{-}}+\frac{k_{B}T}{\Lambda_{0}^3}e^{\beta\bar{\mu}_{0}},
\end{equation}
we obtain the well-known Poisson-Boltzmann-Langevin (PBL) equation \cite{abrashkin2007dipolar,coalson1996statistical}
\begin{equation}
\label{PBL_eq}
-\nabla\cdot(\epsilon(\bold{r})\nabla\psi(\bold{r}))=q_{+}c_{+,b}e^{-\beta q_{+}\psi(\bold{r})}+q_{-}c_{-,b}e^{-\beta q_{-}\psi(\bold{r})},
\end{equation}
where 
\begin{equation}
\epsilon(\bold{r})=\varepsilon_{b} + \frac{c_{0,b} p^2}{k_{B}T} \mathcal{G}(\beta p\mathcal{E}(\bold{r})),
\end{equation}
$\mathcal{G}(x)=\sinh x L(x)/x^2$, $c_{0,b}$ is the bulk concentration of the molecules. Note that we took into consideration the relations for bulk chemical potentials $\mu_{\pm}=k_{B}T\ln(c_{\pm,b}\Lambda_{\pm}^3)$, $\mu_{0}=k_{B}T\ln(c_{0,b}\Lambda_{0}^3)$. To take into account the excluded volume of the ions and molecules, we can use a three-component hard-sphere system as the reference system in the Percus-Yevick approximation determined by eqs. (\ref{PY}-\ref{mu}), where 
$n_0 = \bar{c}_{+} + \bar{c}_{-} + \bar{c}_{0}$, $n_1 = 1/2 ~ d_+ \bar{c}_{+} + 1/2 ~ d_- \bar{c}_{-}+1/2 ~ d_0 \bar{c}_{0}$, $n_2 = \pi~d_+^2 \bar{c}_{+} +\pi ~ d_-^2 \bar{c}_{-}+\pi ~ d_0^2 \bar{c}_{0}$, $n_3 = \pi/6 ~ d_+^3 \bar{c}_{+} + \pi/6 ~ d_-^3 \bar{c}_{-}+\pi/6 ~ d_0^3 \bar{c}_{0}$ ($d_0$ is the effective diameter of molecules).

\section{Stress tensor and macroscopic forces in ionic fluids}
As it has been shown above, the GTP of an ionic fluid for all considered models in the mean-field approximation can be written as follows
\begin{equation}
\label{omega2}
\Omega=\int d\bold{r}\omega(\psi,\nabla\psi),
\end{equation}
where we have introduced the GTP density
\begin{equation}
\omega=-\frac{\varepsilon_{b}(\nabla{\psi})^2}{2}-P.
\end{equation}
We assume that the dependence of the GTP density on coordinates stems only from the electrostatic potential $\psi(\bold{r})$, so that for simplicity we consider the case $u_{\pm}(\bold{r})=0$. The case $u_{\pm}(\bold{r})\neq 0$ will be considered below. 
Variation of the GTP leads to self-consistent field equation (\ref{Poisson_eq_3}), which can be written as follows
\begin{equation}
\label{scf_eq}
\frac{\partial}{\partial x_{i}}\left(\frac{\partial\omega}{\partial\psi_{,i}}\right)-\frac{\partial\omega}{\partial\psi}=0,
\end{equation}
where $x_{i}$ are the Cartesian coordinates and $\psi_{,i}=\partial{\psi}/\partial{x}_{i}$.  Note that we use the Einstein rule, implying the summation over repeated indices. Then, with using eq. (\ref{scf_eq}), we get
\begin{equation}
\frac{\partial\omega}{\partial x_{i}}=\frac{\partial}{\partial x_{k}}\left(\frac{\partial\omega}{\partial\psi_{,k}}\right)\psi_{,i}+\left(\frac{\partial\omega}{\partial\psi_{,k}}\right)\frac{\partial \psi_{,k}}{\partial x_{i}}=\frac{\partial}{\partial x_{k}}\left(\psi_{,i}\frac{\partial{\omega}}{\partial{\psi_{,k}}}\right)
\end{equation}
or 
\begin{equation}
\label{equil_cond}
\frac{\partial \sigma_{ik}}{\partial{x_{k}}}=0,
\end{equation}
where
\begin{equation}
\label{sigma}
\sigma_{ik}=\delta_{ik}\omega-\psi_{,i}\frac{\partial{\omega}}{\partial{\psi_{,k}}}.
\end{equation}
Eq. (\ref{equil_cond}) determines the local mechanical equilibrium condition for the ionic fluid with stress tensor, $\sigma_{ik}$. For model I, eq.(\ref{sigma}) yields
\begin{equation}
\label{stress1}
\sigma_{ik}=\varepsilon_{b}\left(\mathcal{E}_{i}\mathcal{E}_{k}-\frac{1}{2}\mathcal{E}^2\delta_{ik}\right)-\delta_{ik}P_{0},
\end{equation}
where the first term is the standard Maxwell stress tensor \cite{landau2013electrodynamics} in continuous dielectric media with the electrostatic field components $\mathcal{E}_{i}=-\psi_{,i}$, whereas the second term is the stress tensor of an ionic fluid with the local osmotic pressure $P_{0}(\bold{r})=P(T,\bar{\mu}_{+}(\bold{r}),\bar{\mu}_{-}(\bold{r}))$, $\delta_{ik}$ is the unit tensor (Kronecker symbol). For model II, based on the GTP (\ref{Omega4}) and intrinsic chemical potentials (\ref{chem_pot}), we obtain the stress tensor with an account for the polarizabilities and permanent dipole moments of the ions. Using eq. (\ref{sigma}), we obtain the expression for the stress tensor
\begin{equation}
\label{stress2}
\sigma_{ik}=\epsilon \mathcal{E}_{i}\mathcal{E}_{k}-\delta_{ik}\left(P_{0}+\frac{\varepsilon_{b}\mathcal{E}^2}{2}\right).
\end{equation}
Note that in the case of unpolarizable ions ($p_{\pm}=\gamma_{\pm}=0$, $\epsilon=\varepsilon_b$), eq. (\ref{stress2}) transforms into eq. (\ref{stress1}). For model III, using eq. (\ref{sigma}) and taking into account eqs. (\ref{Omega5}), (\ref{chem_pot2}), and (\ref{chem_pot3}), we arrive at the same expression (\ref{stress2}) for the stress tensor with $P_{0}(\bold{r})=P(T,\bar{\mu}_{+}(\bold{r}),\bar{\mu}_{-}(\bold{r}),\bar{\mu}_{0}(\bold{r}))$.

We would like to note that the same approach was used in \cite{landau2013classical} to express the energy-momentum tensor of an arbitrary physical system through its Lagrangian (Noether's theorem~\cite{bogolyubov1973introduction}).
Note that even if we cannot obtain the GTP density from the first principles, we still can utilise eq. (\ref{sigma}) to formulate the stress tensor. However, it would require {\it ad hoc} assumptions regarding the GTP functional dependence  on the electrostatic potential and field.

Now let us consider the case when ions are under the influence of external potential forces with total potential energies $u_{\pm}(\bold{r})$. For example, these can be van der Waals or specific interactions between the ions and the charged surfaces (electrodes or membranes, for instance). In this case, the GTP density depends explicitly on the coordinates via the aforementioned potential energies. Analogous calculations result in the following equation
\begin{equation}
\label{equil_cond_2}
\frac{\partial \sigma_{ik}}{\partial{x_{k}}}-\sum\limits_{\alpha}\bar{c}_{\alpha}\frac{\partial u_{\alpha}}{\partial x_{i}}=0,
\end{equation}
which determines the mechanical equilibrium condition of the ionic fluid in the presence of external volume forces.

Knowledge of the stress tensor allows us to calculate the macroscopic force acting on a conductive (electrode) or dielectric (a membrane or a colloid particle) body immersed in an ionic fluid. It can be calculated as the following integral over the surface of an immersed body
\begin{equation}
\label{force}
F_{i}= \oint\limits_{S} \sigma_{ik}n_{k} dS,
\end{equation}
where $n_k$ are the components of the external normal and $dS$ is the elementary area. Note that in the absence of volume nonelectrostatic forces ($u_{\pm}(\bold{r})=0$), the integration in eq. (\ref{force}) can be performed over any closed surface around a macroscopic body because the ionic fluid is in mechanical equilibrium~\cite{landau2013electrodynamics}. In the presence of nonelectrostatic volume forces, as it follows from the mechanical equilibrium, the macroscopic force can be calculated as follows
\begin{equation}
\label{force2}
F_{i}=\oint\limits_{S} \sigma_{ik}n_{k} dS +\int\limits_{V} f_{i}dV,
\end{equation}
where the first integral is taken over an arbitrary closed surface $S$ around the macroscopic body, whereas the second integral -- over the volume between surface $S$ and the surface of the macroscopic body; $f_{i}=-\sum_{\alpha}\bar{c}_{\alpha}{\partial u_{\alpha}}/{\partial x_{i}}$ are the components of the total nonelectrostatic volume density force. In order to calculate the macroscopic force by eq. (\ref{force}) or (\ref{force2}), it is necessary to solve self-consistent field equation (\ref{Poisson}) with appropriate boundary conditions \footnote{The boundary conditions are determined by the nature of immersed body. The boundary conditions for dielectric or conductive body are standard~\cite{landau2013electrodynamics}. The boundary conditions for surfaces with regulated charge are proposed were~\cite{podgornik2018general}.} to obtain the local osmotic pressure, $P_{0}(\bold{r})$, and local ionic concentrations, $\bar{c}_{\alpha}(\bold{r})$, of the ions and the electrostatic field on the body surface and around it. We would like to note that similar theories based on {\it ad hoc} assumptions were applied in \cite{neu1999wall,trizac1999long,filippov2006electrostatic} to describe the electrostatic interactions between charged particles in ionic fluid within the ideal gas and symmetric lattice gas models. We applied our theory to describe the electrosorption-iduced deformation of electrified porous carbon with nonconvex pores in electrolyte solutions~\cite{kolesnikov2022}.

Now it is instructive to discuss how the general mean-field approximation of disjoining pressure for an ionic liquid in a slit nanopore can be derived. Placing the origin of $z$ axis on one charged wall and another one at $z=H$, we can calculate the disjoining pressure~\cite{derjaguin1987derjaguin} as follows 
\begin{equation}
\Pi=-\frac{\partial\Omega}{\partial H}-P_b,
\end{equation}
where $P_{b}=P(T,\mu_+,\mu_-,\mu_0)$ is the ionic fluid pressure in the bulk. Then, we have
\begin{equation}
\nonumber
\frac{\partial\Omega}{\partial H}=\int\limits_{0}^{H} dz\frac{\delta\Omega}{\delta\psi(z)}\frac{\partial\psi(z)}{\partial H}+\omega-\psi^{\prime}\frac{\partial\omega}{\partial\psi^{\prime}}\bigg|_{z=H}- \int\limits_{0}^{H}dz\sum\limits_{\alpha}\frac{\partial P}{\partial u_{\alpha}}\frac{\partial u_{\alpha}}{\partial H}
\end{equation}
\begin{equation}
= \int\limits_{0}^{H}dz\sum\limits_{\alpha}\bar{c}_{\alpha}(z)\frac{\partial u_{\alpha}(z)}{\partial H}+\sigma_{zz}(H),
\end{equation}
where we took into account that ${\partial P}/{\partial u_{\alpha}}=-{\partial P}/{\partial \bar{\mu}_{\alpha}}=-\bar{c}_{\alpha}$. Therefore, we eventually obtain the general expression for disjoining pressure
\begin{equation}
\Pi=-\sum\limits_{\alpha}\int\limits_{0}^{H}dz\bar{c}_{\alpha}(z)\frac{\partial u_{\alpha}(z)}{\partial H}-\sigma_{zz}(H)-P_b,
\end{equation}
where $\sigma_{zz}(H)=\omega-\psi^{\prime}{\partial\omega}/{\partial\psi^{\prime}}|_{z=H}={\varepsilon_b\mathcal{E}^2(H)}/{2}-P_0(H)$ is the normal stress at $z=H$.
Now let us assume that external potentials are created by identical walls, i.e. 
\begin{equation}
\label{u}
u_{\alpha}(z)=\phi_{\alpha}(z)+\phi_{\alpha}(H-z),
\end{equation}
where $\phi_{\alpha}$ is the single wall potential, and taking into account that $\bar{c}_{\alpha}(z)=\bar{c}_{\alpha}(H-z)$, we arrive at
\begin{equation}
\label{Pi}
\Pi=-\sum\limits_{\alpha}\int\limits_{0}^{H}dz\bar{c}_{\alpha}(z)\phi_{\alpha}^{\prime}(z)-\sigma_{zz}(H)-P_b.
\end{equation}
Eq. (\ref{Pi}) can be rewritten in a form that is more suitable for applications. Using the mechanical equilibrium equation
\begin{equation}
\frac{d\sigma_{zz}}{dz}-\sum\limits_{\alpha}\bar{c}_{\alpha}(z)u_{\alpha}^{\prime}(z)=0,
\end{equation}
after the integration from $z=H/2$ to $z=H$ we obtain
\begin{equation}
\label{sigmazz}
\sigma_{zz}(H)=\sigma_{zz}\left(\frac{H}{2}\right)+\sum\limits_{\alpha}\int\limits_{H/2}^{H}dz\bar{c}_{\alpha}(z)\phi_{\alpha}^{\prime}(z)-\sum\limits_{\alpha}\int\limits_{0}^{H/2}dz\bar{c}_{\alpha}(z)\phi_{\alpha}^{\prime}(z),
\end{equation}
where we have taken into account eq. (\ref{u}) and condition $\bar{c}_{\alpha}(z)=\bar{c}_{\alpha}(H-z)$.
Substituting expression (\ref{sigmazz}) for $\sigma_{zz}(H)$ in eq.(\ref{Pi}), after some algebra, we obtain
\begin{equation}
\label{Pi2}
\Pi=P_{m}-P_{b}-2\sum\limits_{\alpha}\int\limits_{H/2}^{H}dz\bar{c}_{\alpha}(z)\phi_{\alpha}^{\prime}(z),
\end{equation}
where $P_{m}=P_{0}\left({H}/{2}\right)$ is the pressure in the middle of a pore and we take into account that $\mathcal{E}(H/2)=0$. Note that a similar expression for disjoining pressure for simple (nonionic) fluids was obtained by Henderson in Ref.\cite{henderson1986compressibility} within a different approach. For negligible volume forces ($\phi_{\alpha}(z)=0$), we obtain the Langmuir equation~\cite{derjaguin1987derjaguin}, $\Pi=P_m - P_b$, for disjoining pressure. Eq. (\ref{Pi2}) is a generalization of the classical DLVO expression for disjoining pressure for the case of an arbitrary reference fluid model in the presence of external volume forces.

\section{Concluding remarks}
In conclusion, we proposed a field-theoretical approach based on the thermodynamic perturbation theory allowing us to derive the grand thermodynamic potential of the inhomogeneous ionic fluid as a functional of electrostatic potential for an arbitrary reference system. We obtained the modified Poisson-Boltzmann equation as the Euler-Lagrange equation for the grand thermodynamic potential. Using Noether's theorem, we derived a general mean-field expression for the stress tensor consistent with the respective modified Poisson-Boltzmann equation. We derived a general mean-field expression for the disjoining pressure of an ionic fluid in a slit pore. We applied the developed formalism to describe three ionic liquid models of practical importance: nonpolarizable models (including the Poisson-Boltzmann and Poisson-Fermi equations), polarizable models (the ions have nonzero permanent dipole moment or static polarizability), and ion-dipole mixtures (mixture of polar/polarizable ions and polar/polarizable molecules). For all these models, we derived modified Poisson-Boltzmann equations and respective stress tensors, which can be used in different applications, where it is necessary to estimate the electrostatic potential and ionic concentrations around dielectric or conductive bodies (electrodes, colloids, membranes, dust particles, {\it etc}.) immersed in an ionic fluid, as well as the macroscopic forces acting on them.

\section*{Acknowledgments}
The development of the self-consistent field theories presented in section 2 is supported by the Russian Science Foundation (No. 21-11-00031). Theory presented in section 3 is supported by grant of the President of the Russian Federation (project No. MD-341.2021.1.3).
\bibliography{name}

\begin{thebibliography}{44}%
\makeatletter
\providecommand \@ifxundefined [1]{%
 \@ifx{#1\undefined}
}%
\providecommand \@ifnum [1]{%
 \ifnum #1\expandafter \@firstoftwo
 \else \expandafter \@secondoftwo
 \fi
}%
\providecommand \@ifx [1]{%
 \ifx #1\expandafter \@firstoftwo
 \else \expandafter \@secondoftwo
 \fi
}%
\providecommand \natexlab [1]{#1}%
\providecommand \enquote  [1]{``#1''}%
\providecommand \bibnamefont  [1]{#1}%
\providecommand \bibfnamefont [1]{#1}%
\providecommand \citenamefont [1]{#1}%
\providecommand \href@noop [0]{\@secondoftwo}%
\providecommand \href [0]{\begingroup \@sanitize@url \@href}%
\providecommand \@href[1]{\@@startlink{#1}\@@href}%
\providecommand \@@href[1]{\endgroup#1\@@endlink}%
\providecommand \@sanitize@url [0]{\catcode `\\12\catcode `\$12\catcode
  `\&12\catcode `\#12\catcode `\^12\catcode `\_12\catcode `\%12\relax}%
\providecommand \@@startlink[1]{}%
\providecommand \@@endlink[0]{}%
\providecommand \url  [0]{\begingroup\@sanitize@url \@url }%
\providecommand \@url [1]{\endgroup\@href {#1}{\urlprefix }}%
\providecommand \urlprefix  [0]{URL }%
\providecommand \Eprint [0]{\href }%
\providecommand \doibase [0]{http://dx.doi.org/}%
\providecommand \selectlanguage [0]{\@gobble}%
\providecommand \bibinfo  [0]{\@secondoftwo}%
\providecommand \bibfield  [0]{\@secondoftwo}%
\providecommand \translation [1]{[#1]}%
\providecommand \BibitemOpen [0]{}%
\providecommand \bibitemStop [0]{}%
\providecommand \bibitemNoStop [0]{.\EOS\space}%
\providecommand \EOS [0]{\spacefactor3000\relax}%
\providecommand \BibitemShut  [1]{\csname bibitem#1\endcsname}%
\let\auto@bib@innerbib\@empty
\bibitem [{\citenamefont {Kralj-Igli{\v{c}}}\ and\ \citenamefont
  {Igli{\v{c}}}(1996)}]{kralj1996simple}%
  \BibitemOpen
  \bibfield  {author} {\bibinfo {author} {\bibfnamefont {V.}~\bibnamefont
  {Kralj-Igli{\v{c}}}}\ and\ \bibinfo {author} {\bibfnamefont {A.}~\bibnamefont
  {Igli{\v{c}}}},\ }\bibfield  {title} {\enquote {\bibinfo {title} {A simple
  statistical mechanical approach to the free energy of the electric double
  layer including the excluded volume effect},}\ }\href@noop {} {\bibfield
  {journal} {\bibinfo  {journal} {Journal de Physique II}\ }\textbf {\bibinfo
  {volume} {6}},\ \bibinfo {pages} {477--491} (\bibinfo {year}
  {1996})}\BibitemShut {NoStop}%
\bibitem [{\citenamefont {Borukhov}, \citenamefont {Andelman},\ and\
  \citenamefont {Orland}(1997)}]{borukhov1997steric}%
  \BibitemOpen
  \bibfield  {author} {\bibinfo {author} {\bibfnamefont {I.}~\bibnamefont
  {Borukhov}}, \bibinfo {author} {\bibfnamefont {D.}~\bibnamefont {Andelman}},
  \ and\ \bibinfo {author} {\bibfnamefont {H.}~\bibnamefont {Orland}},\
  }\bibfield  {title} {\enquote {\bibinfo {title} {Steric effects in
  electrolytes: A modified poisson-boltzmann equation},}\ }\href@noop {}
  {\bibfield  {journal} {\bibinfo  {journal} {Physical review letters}\
  }\textbf {\bibinfo {volume} {79}},\ \bibinfo {pages} {435} (\bibinfo {year}
  {1997})}\BibitemShut {NoStop}%
\bibitem [{\citenamefont {Maggs}\ and\ \citenamefont
  {Podgornik}(2016)}]{maggs2016general}%
  \BibitemOpen
  \bibfield  {author} {\bibinfo {author} {\bibfnamefont {A.}~\bibnamefont
  {Maggs}}\ and\ \bibinfo {author} {\bibfnamefont {R.}~\bibnamefont
  {Podgornik}},\ }\bibfield  {title} {\enquote {\bibinfo {title} {General
  theory of asymmetric steric interactions in electrostatic double layers},}\
  }\href@noop {} {\bibfield  {journal} {\bibinfo  {journal} {Soft matter}\
  }\textbf {\bibinfo {volume} {12}},\ \bibinfo {pages} {1219--1229} (\bibinfo
  {year} {2016})}\BibitemShut {NoStop}%
\bibitem [{\citenamefont {Kornyshev}(2007)}]{kornyshev2007double}%
  \BibitemOpen
  \bibfield  {author} {\bibinfo {author} {\bibfnamefont {A.~A.}\ \bibnamefont
  {Kornyshev}},\ }\bibfield  {title} {\enquote {\bibinfo {title} {Double-layer
  in ionic liquids: paradigm change?}}\ }\href@noop {} {\bibfield  {journal}
  {\bibinfo  {journal} {The Journal of Physical Chemistry B}\ }\textbf
  {\bibinfo {volume} {111}},\ \bibinfo {pages} {5545--5557} (\bibinfo {year}
  {2007})}\BibitemShut {NoStop}%
\bibitem [{\citenamefont {Goodwin}, \citenamefont {Feng},\ and\ \citenamefont
  {Kornyshev}(2017)}]{goodwin2017mean}%
  \BibitemOpen
  \bibfield  {author} {\bibinfo {author} {\bibfnamefont {Z.~A.}\ \bibnamefont
  {Goodwin}}, \bibinfo {author} {\bibfnamefont {G.}~\bibnamefont {Feng}}, \
  and\ \bibinfo {author} {\bibfnamefont {A.~A.}\ \bibnamefont {Kornyshev}},\
  }\bibfield  {title} {\enquote {\bibinfo {title} {Mean-field theory of
  electrical double layer in ionic liquids with account of short-range
  correlations},}\ }\href@noop {} {\bibfield  {journal} {\bibinfo  {journal}
  {Electrochimica Acta}\ }\textbf {\bibinfo {volume} {225}},\ \bibinfo {pages}
  {190--197} (\bibinfo {year} {2017})}\BibitemShut {NoStop}%
\bibitem [{\citenamefont {Blossey}, \citenamefont {Maggs},\ and\ \citenamefont
  {Podgornik}(2017)}]{blossey2017structural}%
  \BibitemOpen
  \bibfield  {author} {\bibinfo {author} {\bibfnamefont {R.}~\bibnamefont
  {Blossey}}, \bibinfo {author} {\bibfnamefont {A.}~\bibnamefont {Maggs}}, \
  and\ \bibinfo {author} {\bibfnamefont {R.}~\bibnamefont {Podgornik}},\
  }\bibfield  {title} {\enquote {\bibinfo {title} {Structural interactions in
  ionic liquids linked to higher-order poisson-boltzmann equations},}\
  }\href@noop {} {\bibfield  {journal} {\bibinfo  {journal} {Physical Review
  E}\ }\textbf {\bibinfo {volume} {95}},\ \bibinfo {pages} {060602} (\bibinfo
  {year} {2017})}\BibitemShut {NoStop}%
\bibitem [{\citenamefont {Frydel}(2011)}]{frydel2011polarizable}%
  \BibitemOpen
  \bibfield  {author} {\bibinfo {author} {\bibfnamefont {D.}~\bibnamefont
  {Frydel}},\ }\bibfield  {title} {\enquote {\bibinfo {title} {Polarizable
  poisson--boltzmann equation: The study of polarizability effects on the
  structure of a double layer},}\ }\href@noop {} {\bibfield  {journal}
  {\bibinfo  {journal} {The Journal of chemical physics}\ }\textbf {\bibinfo
  {volume} {134}},\ \bibinfo {pages} {234704} (\bibinfo {year}
  {2011})}\BibitemShut {NoStop}%
\bibitem [{\citenamefont {Budkov}\ \emph {et~al.}(2020)\citenamefont {Budkov},
  \citenamefont {Sergeev}, \citenamefont {Zavarzin},\ and\ \citenamefont
  {Kolesnikov}}]{budkov2020two}%
  \BibitemOpen
  \bibfield  {author} {\bibinfo {author} {\bibfnamefont {Y.~A.}\ \bibnamefont
  {Budkov}}, \bibinfo {author} {\bibfnamefont {A.~V.}\ \bibnamefont {Sergeev}},
  \bibinfo {author} {\bibfnamefont {S.~V.}\ \bibnamefont {Zavarzin}}, \ and\
  \bibinfo {author} {\bibfnamefont {A.~L.}\ \bibnamefont {Kolesnikov}},\
  }\bibfield  {title} {\enquote {\bibinfo {title} {Two-component electrolyte
  solutions with dipolar cations on a charged electrode: Theory and computer
  simulations},}\ }\href@noop {} {\bibfield  {journal} {\bibinfo  {journal}
  {The Journal of Physical Chemistry C}\ }\textbf {\bibinfo {volume} {124}},\
  \bibinfo {pages} {16308--16314} (\bibinfo {year} {2020})}\BibitemShut
  {NoStop}%
\bibitem [{\citenamefont {Budkov}, \citenamefont {Zavarzin},\ and\
  \citenamefont {Kolesnikov}(2021)}]{budkov2021theory}%
  \BibitemOpen
  \bibfield  {author} {\bibinfo {author} {\bibfnamefont {Y.~A.}\ \bibnamefont
  {Budkov}}, \bibinfo {author} {\bibfnamefont {S.~V.}\ \bibnamefont
  {Zavarzin}}, \ and\ \bibinfo {author} {\bibfnamefont {A.~L.}\ \bibnamefont
  {Kolesnikov}},\ }\bibfield  {title} {\enquote {\bibinfo {title} {Theory of
  ionic liquids with polarizable ions on a charged electrode},}\ }\href@noop {}
  {\bibfield  {journal} {\bibinfo  {journal} {The Journal of Physical Chemistry
  C}\ }\textbf {\bibinfo {volume} {125}},\ \bibinfo {pages} {21151--21159}
  (\bibinfo {year} {2021})}\BibitemShut {NoStop}%
\bibitem [{\citenamefont {Budkov}, \citenamefont {Kolesnikov},\ and\
  \citenamefont {Kiselev}(2015)}]{budkov2015modified}%
  \BibitemOpen
  \bibfield  {author} {\bibinfo {author} {\bibfnamefont {Y.~A.}\ \bibnamefont
  {Budkov}}, \bibinfo {author} {\bibfnamefont {A.}~\bibnamefont {Kolesnikov}},
  \ and\ \bibinfo {author} {\bibfnamefont {M.}~\bibnamefont {Kiselev}},\
  }\bibfield  {title} {\enquote {\bibinfo {title} {A modified poisson-boltzmann
  theory: Effects of co-solvent polarizability},}\ }\href@noop {} {\bibfield
  {journal} {\bibinfo  {journal} {EPL (Europhysics Letters)}\ }\textbf
  {\bibinfo {volume} {111}},\ \bibinfo {pages} {28002} (\bibinfo {year}
  {2015})}\BibitemShut {NoStop}%
\bibitem [{\citenamefont {Budkov}, \citenamefont {Kolesnikov},\ and\
  \citenamefont {Kiselev}(2016)}]{budkov2016theory}%
  \BibitemOpen
  \bibfield  {author} {\bibinfo {author} {\bibfnamefont {Y.~A.}\ \bibnamefont
  {Budkov}}, \bibinfo {author} {\bibfnamefont {A.}~\bibnamefont {Kolesnikov}},
  \ and\ \bibinfo {author} {\bibfnamefont {M.}~\bibnamefont {Kiselev}},\
  }\bibfield  {title} {\enquote {\bibinfo {title} {On the theory of electric
  double layer with explicit account of a polarizable co-solvent},}\
  }\href@noop {} {\bibfield  {journal} {\bibinfo  {journal} {The Journal of
  Chemical Physics}\ }\textbf {\bibinfo {volume} {144}},\ \bibinfo {pages}
  {184703} (\bibinfo {year} {2016})}\BibitemShut {NoStop}%
\bibitem [{\citenamefont {Budkov}\ \emph {et~al.}(2018)\citenamefont {Budkov},
  \citenamefont {Kolesnikov}, \citenamefont {Goodwin}, \citenamefont
  {Kiselev},\ and\ \citenamefont {Kornyshev}}]{budkov2018theory}%
  \BibitemOpen
  \bibfield  {author} {\bibinfo {author} {\bibfnamefont {Y.~A.}\ \bibnamefont
  {Budkov}}, \bibinfo {author} {\bibfnamefont {A.~L.}\ \bibnamefont
  {Kolesnikov}}, \bibinfo {author} {\bibfnamefont {Z.~A.}\ \bibnamefont
  {Goodwin}}, \bibinfo {author} {\bibfnamefont {M.~G.}\ \bibnamefont
  {Kiselev}}, \ and\ \bibinfo {author} {\bibfnamefont {A.~A.}\ \bibnamefont
  {Kornyshev}},\ }\bibfield  {title} {\enquote {\bibinfo {title} {Theory of
  electrosorption of water from ionic liquids},}\ }\href@noop {} {\bibfield
  {journal} {\bibinfo  {journal} {Electrochimica Acta}\ }\textbf {\bibinfo
  {volume} {284}},\ \bibinfo {pages} {346--354} (\bibinfo {year}
  {2018})}\BibitemShut {NoStop}%
\bibitem [{\citenamefont {Coalson}\ and\ \citenamefont
  {Duncan}(1996)}]{coalson1996statistical}%
  \BibitemOpen
  \bibfield  {author} {\bibinfo {author} {\bibfnamefont {R.~D.}\ \bibnamefont
  {Coalson}}\ and\ \bibinfo {author} {\bibfnamefont {A.}~\bibnamefont
  {Duncan}},\ }\bibfield  {title} {\enquote {\bibinfo {title} {Statistical
  mechanics of a multipolar gas: a lattice field theory approach},}\
  }\href@noop {} {\bibfield  {journal} {\bibinfo  {journal} {The Journal of
  Physical Chemistry}\ }\textbf {\bibinfo {volume} {100}},\ \bibinfo {pages}
  {2612--2620} (\bibinfo {year} {1996})}\BibitemShut {NoStop}%
\bibitem [{\citenamefont {Abrashkin}, \citenamefont {Andelman},\ and\
  \citenamefont {Orland}(2007)}]{abrashkin2007dipolar}%
  \BibitemOpen
  \bibfield  {author} {\bibinfo {author} {\bibfnamefont {A.}~\bibnamefont
  {Abrashkin}}, \bibinfo {author} {\bibfnamefont {D.}~\bibnamefont {Andelman}},
  \ and\ \bibinfo {author} {\bibfnamefont {H.}~\bibnamefont {Orland}},\
  }\bibfield  {title} {\enquote {\bibinfo {title} {Dipolar poisson-boltzmann
  equation: ions and dipoles close to charge interfaces},}\ }\href@noop {}
  {\bibfield  {journal} {\bibinfo  {journal} {Physical review letters}\
  }\textbf {\bibinfo {volume} {99}},\ \bibinfo {pages} {077801} (\bibinfo
  {year} {2007})}\BibitemShut {NoStop}%
\bibitem [{\citenamefont {Igli{\v{c}}}, \citenamefont {Gongadze},\ and\
  \citenamefont {Bohinc}(2010)}]{iglivc2010excluded}%
  \BibitemOpen
  \bibfield  {author} {\bibinfo {author} {\bibfnamefont {A.}~\bibnamefont
  {Igli{\v{c}}}}, \bibinfo {author} {\bibfnamefont {E.}~\bibnamefont
  {Gongadze}}, \ and\ \bibinfo {author} {\bibfnamefont {K.}~\bibnamefont
  {Bohinc}},\ }\bibfield  {title} {\enquote {\bibinfo {title} {Excluded volume
  effect and orientational ordering near charged surface in solution of ions
  and langevin dipoles},}\ }\href@noop {} {\bibfield  {journal} {\bibinfo
  {journal} {Bioelectrochemistry}\ }\textbf {\bibinfo {volume} {79}},\ \bibinfo
  {pages} {223--227} (\bibinfo {year} {2010})}\BibitemShut {NoStop}%
\bibitem [{\citenamefont {Buyukdagli}\ and\ \citenamefont
  {Ala-Nissila}(2013)}]{buyukdagli2013microscopic}%
  \BibitemOpen
  \bibfield  {author} {\bibinfo {author} {\bibfnamefont {S.}~\bibnamefont
  {Buyukdagli}}\ and\ \bibinfo {author} {\bibfnamefont {T.}~\bibnamefont
  {Ala-Nissila}},\ }\bibfield  {title} {\enquote {\bibinfo {title} {Microscopic
  formulation of nonlocal electrostatics in polar liquids embedding polarizable
  ions},}\ }\href@noop {} {\bibfield  {journal} {\bibinfo  {journal} {Physical
  Review E}\ }\textbf {\bibinfo {volume} {87}},\ \bibinfo {pages} {063201}
  (\bibinfo {year} {2013})}\BibitemShut {NoStop}%
\bibitem [{\citenamefont {Slavchov}(2014)}]{slavchov2014quadrupole}%
  \BibitemOpen
  \bibfield  {author} {\bibinfo {author} {\bibfnamefont {R.~I.}\ \bibnamefont
  {Slavchov}},\ }\bibfield  {title} {\enquote {\bibinfo {title} {Quadrupole
  terms in the maxwell equations: Debye-h{\"u}ckel theory in quadrupolarizable
  solvent and self-salting-out of electrolytes},}\ }\href@noop {} {\bibfield
  {journal} {\bibinfo  {journal} {The Journal of Chemical Physics}\ }\textbf
  {\bibinfo {volume} {140}},\ \bibinfo {pages} {074503} (\bibinfo {year}
  {2014})}\BibitemShut {NoStop}%
\bibitem [{\citenamefont {Podgornik}(2018)}]{podgornik2018general}%
  \BibitemOpen
  \bibfield  {author} {\bibinfo {author} {\bibfnamefont {R.}~\bibnamefont
  {Podgornik}},\ }\bibfield  {title} {\enquote {\bibinfo {title} {General
  theory of charge regulation and surface differential capacitance},}\
  }\href@noop {} {\bibfield  {journal} {\bibinfo  {journal} {The Journal of
  chemical physics}\ }\textbf {\bibinfo {volume} {149}},\ \bibinfo {pages}
  {104701} (\bibinfo {year} {2018})}\BibitemShut {NoStop}%
\bibitem [{\citenamefont {Uematsu}, \citenamefont {Netz},\ and\ \citenamefont
  {Bonthuis}(2018)}]{uematsu2018effects}%
  \BibitemOpen
  \bibfield  {author} {\bibinfo {author} {\bibfnamefont {Y.}~\bibnamefont
  {Uematsu}}, \bibinfo {author} {\bibfnamefont {R.~R.}\ \bibnamefont {Netz}}, \
  and\ \bibinfo {author} {\bibfnamefont {D.~J.}\ \bibnamefont {Bonthuis}},\
  }\bibfield  {title} {\enquote {\bibinfo {title} {The effects of ion
  adsorption on the potential of zero charge and the differential capacitance
  of charged aqueous interfaces},}\ }\href@noop {} {\bibfield  {journal}
  {\bibinfo  {journal} {Journal of Physics: Condensed Matter}\ }\textbf
  {\bibinfo {volume} {30}},\ \bibinfo {pages} {064002} (\bibinfo {year}
  {2018})}\BibitemShut {NoStop}%
\bibitem [{\citenamefont {Buyukdagli}(2020)}]{buyukdagli2020schwinger}%
  \BibitemOpen
  \bibfield  {author} {\bibinfo {author} {\bibfnamefont {S.}~\bibnamefont
  {Buyukdagli}},\ }\bibfield  {title} {\enquote {\bibinfo {title}
  {Schwinger-dyson equations for composite electrolytes governed by mixed
  electrostatic couplings strengths},}\ }\href@noop {} {\bibfield  {journal}
  {\bibinfo  {journal} {The Journal of chemical physics}\ }\textbf {\bibinfo
  {volume} {152}},\ \bibinfo {pages} {014902} (\bibinfo {year}
  {2020})}\BibitemShut {NoStop}%
\bibitem [{\citenamefont {de~Souza}\ and\ \citenamefont
  {Bazant}(2020)}]{de2020continuum}%
  \BibitemOpen
  \bibfield  {author} {\bibinfo {author} {\bibfnamefont {J.~P.}\ \bibnamefont
  {de~Souza}}\ and\ \bibinfo {author} {\bibfnamefont {M.~Z.}\ \bibnamefont
  {Bazant}},\ }\bibfield  {title} {\enquote {\bibinfo {title} {Continuum theory
  of electrostatic correlations at charged surfaces},}\ }\href@noop {}
  {\bibfield  {journal} {\bibinfo  {journal} {The Journal of Physical Chemistry
  C}\ }\textbf {\bibinfo {volume} {124}},\ \bibinfo {pages} {11414--11421}
  (\bibinfo {year} {2020})}\BibitemShut {NoStop}%
\bibitem [{\citenamefont {Bazant}, \citenamefont {Storey},\ and\ \citenamefont
  {Kornyshev}(2011)}]{bazant2011double}%
  \BibitemOpen
  \bibfield  {author} {\bibinfo {author} {\bibfnamefont {M.~Z.}\ \bibnamefont
  {Bazant}}, \bibinfo {author} {\bibfnamefont {B.~D.}\ \bibnamefont {Storey}},
  \ and\ \bibinfo {author} {\bibfnamefont {A.~A.}\ \bibnamefont {Kornyshev}},\
  }\bibfield  {title} {\enquote {\bibinfo {title} {Double layer in ionic
  liquids: Overscreening versus crowding},}\ }\href@noop {} {\bibfield
  {journal} {\bibinfo  {journal} {Physical review letters}\ }\textbf {\bibinfo
  {volume} {106}},\ \bibinfo {pages} {046102} (\bibinfo {year}
  {2011})}\BibitemShut {NoStop}%
\bibitem [{\citenamefont {Kubo}(1962)}]{kubo1962generalized}%
  \BibitemOpen
  \bibfield  {author} {\bibinfo {author} {\bibfnamefont {R.}~\bibnamefont
  {Kubo}},\ }\bibfield  {title} {\enquote {\bibinfo {title} {Generalized
  cumulant expansion method},}\ }\href@noop {} {\bibfield  {journal} {\bibinfo
  {journal} {Journal of the Physical Society of Japan}\ }\textbf {\bibinfo
  {volume} {17}},\ \bibinfo {pages} {1100--1120} (\bibinfo {year}
  {1962})}\BibitemShut {NoStop}%
\bibitem [{\citenamefont {Zinn-Justin}(2002)}]{zinn2002quantum}%
  \BibitemOpen
  \bibfield  {author} {\bibinfo {author} {\bibfnamefont {J.}~\bibnamefont
  {Zinn-Justin}},\ }\href@noop {} {\emph {\bibinfo {title} {Quantum field
  theory and critical phenomena}}},\ Vol.\ \bibinfo {volume} {113}\ (\bibinfo
  {publisher} {Clarendon Press, Oxford},\ \bibinfo {year} {2002})\BibitemShut
  {NoStop}%
\bibitem [{\citenamefont {Hill}(1986)}]{hill1986introduction}%
  \BibitemOpen
  \bibfield  {author} {\bibinfo {author} {\bibfnamefont {T.~L.}\ \bibnamefont
  {Hill}},\ }\href@noop {} {\emph {\bibinfo {title} {An introduction to
  statistical thermodynamics}}}\ (\bibinfo  {publisher} {Courier Corporation},\
  \bibinfo {year} {1986})\BibitemShut {NoStop}%
\bibitem [{\citenamefont {Ivanchenko}\ and\ \citenamefont
  {Lisyanskii}(1984)}]{ivanchenko1984ginzburg}%
  \BibitemOpen
  \bibfield  {author} {\bibinfo {author} {\bibfnamefont {Y.~M.}\ \bibnamefont
  {Ivanchenko}}\ and\ \bibinfo {author} {\bibfnamefont {A.}~\bibnamefont
  {Lisyanskii}},\ }\bibfield  {title} {\enquote {\bibinfo {title}
  {Ginzburg-landau functional for liquid-vapor phase transition},}\ }\href@noop
  {} {\bibfield  {journal} {\bibinfo  {journal} {Theoretical and Mathematical
  Physics}\ }\textbf {\bibinfo {volume} {58}},\ \bibinfo {pages} {97--103}
  (\bibinfo {year} {1984})}\BibitemShut {NoStop}%
\bibitem [{\citenamefont {Brilliantov}(1998)}]{brilliantov1998effective}%
  \BibitemOpen
  \bibfield  {author} {\bibinfo {author} {\bibfnamefont {N.~V.}\ \bibnamefont
  {Brilliantov}},\ }\bibfield  {title} {\enquote {\bibinfo {title} {Effective
  magnetic hamiltonian and ginzburg criterion for fluids},}\ }\href@noop {}
  {\bibfield  {journal} {\bibinfo  {journal} {Physical Review E}\ }\textbf
  {\bibinfo {volume} {58}},\ \bibinfo {pages} {2628} (\bibinfo {year}
  {1998})}\BibitemShut {NoStop}%
\bibitem [{\citenamefont {Brilliantov}, \citenamefont {Rub{\'\i}},\ and\
  \citenamefont {Budkov}(2020)}]{brilliantov2020molecular}%
  \BibitemOpen
  \bibfield  {author} {\bibinfo {author} {\bibfnamefont {N.~V.}\ \bibnamefont
  {Brilliantov}}, \bibinfo {author} {\bibfnamefont {J.~M.}\ \bibnamefont
  {Rub{\'\i}}}, \ and\ \bibinfo {author} {\bibfnamefont {Y.~A.}\ \bibnamefont
  {Budkov}},\ }\bibfield  {title} {\enquote {\bibinfo {title} {Molecular fields
  and statistical field theory of fluids: Application to interface
  phenomena},}\ }\href@noop {} {\bibfield  {journal} {\bibinfo  {journal}
  {Physical Review E}\ }\textbf {\bibinfo {volume} {101}},\ \bibinfo {pages}
  {042135} (\bibinfo {year} {2020})}\BibitemShut {NoStop}%
\bibitem [{\citenamefont {Bazant}\ \emph {et~al.}(2009)\citenamefont {Bazant},
  \citenamefont {Kilic}, \citenamefont {Storey},\ and\ \citenamefont
  {Ajdari}}]{bazant2009towards}%
  \BibitemOpen
  \bibfield  {author} {\bibinfo {author} {\bibfnamefont {M.~Z.}\ \bibnamefont
  {Bazant}}, \bibinfo {author} {\bibfnamefont {M.~S.}\ \bibnamefont {Kilic}},
  \bibinfo {author} {\bibfnamefont {B.~D.}\ \bibnamefont {Storey}}, \ and\
  \bibinfo {author} {\bibfnamefont {A.}~\bibnamefont {Ajdari}},\ }\bibfield
  {title} {\enquote {\bibinfo {title} {Towards an understanding of
  induced-charge electrokinetics at large applied voltages in concentrated
  solutions},}\ }\href@noop {} {\bibfield  {journal} {\bibinfo  {journal}
  {Advances in colloid and interface science}\ }\textbf {\bibinfo {volume}
  {152}},\ \bibinfo {pages} {48--88} (\bibinfo {year} {2009})}\BibitemShut
  {NoStop}%
\bibitem [{\citenamefont {Roth}(2010)}]{roth2010fundamental}%
  \BibitemOpen
  \bibfield  {author} {\bibinfo {author} {\bibfnamefont {R.}~\bibnamefont
  {Roth}},\ }\bibfield  {title} {\enquote {\bibinfo {title} {Fundamental
  measure theory for hard-sphere mixtures: a review},}\ }\href@noop {}
  {\bibfield  {journal} {\bibinfo  {journal} {Journal of Physics: Condensed
  Matter}\ }\textbf {\bibinfo {volume} {22}},\ \bibinfo {pages} {063102}
  (\bibinfo {year} {2010})}\BibitemShut {NoStop}%
\bibitem [{\citenamefont {Roth}\ \emph {et~al.}(2002)\citenamefont {Roth},
  \citenamefont {Evans}, \citenamefont {Lang},\ and\ \citenamefont
  {Kahl}}]{roth2002fundamental}%
  \BibitemOpen
  \bibfield  {author} {\bibinfo {author} {\bibfnamefont {R.}~\bibnamefont
  {Roth}}, \bibinfo {author} {\bibfnamefont {R.}~\bibnamefont {Evans}},
  \bibinfo {author} {\bibfnamefont {A.}~\bibnamefont {Lang}}, \ and\ \bibinfo
  {author} {\bibfnamefont {G.}~\bibnamefont {Kahl}},\ }\bibfield  {title}
  {\enquote {\bibinfo {title} {Fundamental measure theory for hard-sphere
  mixtures revisited: the white bear version},}\ }\href@noop {} {\bibfield
  {journal} {\bibinfo  {journal} {Journal of Physics: Condensed Matter}\
  }\textbf {\bibinfo {volume} {14}},\ \bibinfo {pages} {12063} (\bibinfo {year}
  {2002})}\BibitemShut {NoStop}%
\bibitem [{\citenamefont {Budkov}(2019)}]{budkov2019statistical}%
  \BibitemOpen
  \bibfield  {author} {\bibinfo {author} {\bibfnamefont {Y.~A.}\ \bibnamefont
  {Budkov}},\ }\bibfield  {title} {\enquote {\bibinfo {title} {A statistical
  field theory of salt solutions of "hairy" dielectric particles},}\
  }\href@noop {} {\bibfield  {journal} {\bibinfo  {journal} {Journal of
  Physics: Condensed Matter}\ }\textbf {\bibinfo {volume} {32}},\ \bibinfo
  {pages} {055101} (\bibinfo {year} {2019})}\BibitemShut {NoStop}%
\bibitem [{\citenamefont {Budkov}(2020)}]{budkov2020statistical}%
  \BibitemOpen
  \bibfield  {author} {\bibinfo {author} {\bibfnamefont {Y.~A.}\ \bibnamefont
  {Budkov}},\ }\bibfield  {title} {\enquote {\bibinfo {title} {Statistical
  field theory of ion--molecular solutions},}\ }\href@noop {} {\bibfield
  {journal} {\bibinfo  {journal} {Physical Chemistry Chemical Physics}\
  }\textbf {\bibinfo {volume} {22}},\ \bibinfo {pages} {14756--14772} (\bibinfo
  {year} {2020})}\BibitemShut {NoStop}%
\bibitem [{\citenamefont {Budkov}(2018)}]{budkov2018nonlocal}%
  \BibitemOpen
  \bibfield  {author} {\bibinfo {author} {\bibfnamefont {Y.~A.}\ \bibnamefont
  {Budkov}},\ }\bibfield  {title} {\enquote {\bibinfo {title} {Nonlocal
  statistical field theory of dipolar particles in electrolyte solutions},}\
  }\href@noop {} {\bibfield  {journal} {\bibinfo  {journal} {Journal of
  Physics: Condensed Matter}\ }\textbf {\bibinfo {volume} {30}},\ \bibinfo
  {pages} {344001} (\bibinfo {year} {2018})}\BibitemShut {NoStop}%
\bibitem [{\citenamefont {Landau}, \citenamefont {Lifshitz},\ and\
  \citenamefont {Pitaevskii}(1984)}]{landau2013electrodynamics}%
  \BibitemOpen
  \bibfield  {author} {\bibinfo {author} {\bibfnamefont {L.~D.}\ \bibnamefont
  {Landau}}, \bibinfo {author} {\bibfnamefont {E.~M.}\ \bibnamefont
  {Lifshitz}}, \ and\ \bibinfo {author} {\bibfnamefont {L.~P.}\ \bibnamefont
  {Pitaevskii}},\ }\href@noop {} {\emph {\bibinfo {title} {Course of
  Theoretical Physics VIII: Electrodynamics of continuous media}}}\ (\bibinfo
  {publisher} {Pergamon Press},\ \bibinfo {year} {1984})\BibitemShut {NoStop}%
\bibitem [{\citenamefont {Landau}\ and\ \citenamefont
  {Lifshitz}(2013)}]{landau2013classical}%
  \BibitemOpen
  \bibfield  {author} {\bibinfo {author} {\bibfnamefont {L.~D.}\ \bibnamefont
  {Landau}}\ and\ \bibinfo {author} {\bibfnamefont {E.~M.}\ \bibnamefont
  {Lifshitz}},\ }\href@noop {} {\emph {\bibinfo {title} {The classical theory
  of fields}}}\ (\bibinfo  {publisher} {Elsevier},\ \bibinfo {year}
  {2013})\BibitemShut {NoStop}%
\bibitem [{\citenamefont {Bogolyubov}\ and\ \citenamefont
  {Shirkov}(1973)}]{bogolyubov1973introduction}%
  \BibitemOpen
  \bibfield  {author} {\bibinfo {author} {\bibfnamefont {N.}~\bibnamefont
  {Bogolyubov}}\ and\ \bibinfo {author} {\bibfnamefont {D.}~\bibnamefont
  {Shirkov}},\ }\bibfield  {title} {\enquote {\bibinfo {title} {Introduction to
  quantum fields theory},}\ }\href@noop {} {\bibfield  {journal} {\bibinfo
  {journal} {Nauka Eds}\ } (\bibinfo {year} {1973})}\BibitemShut {NoStop}%
\bibitem [{Note1()}]{Note1}%
  \BibitemOpen
  \bibinfo {note} {The boundary conditions are determined by the nature of
  immersed body. The boundary conditions for dielectric or conductive body are
  standard~\cite {landau2013electrodynamics}. The boundary conditions for
  surfaces with regulated charge are proposed were~\cite
  {podgornik2018general}.}\BibitemShut {Stop}%
\bibitem [{\citenamefont {Neu}(1999)}]{neu1999wall}%
  \BibitemOpen
  \bibfield  {author} {\bibinfo {author} {\bibfnamefont {J.~C.}\ \bibnamefont
  {Neu}},\ }\bibfield  {title} {\enquote {\bibinfo {title} {Wall-mediated
  forces between like-charged bodies in an electrolyte},}\ }\href@noop {}
  {\bibfield  {journal} {\bibinfo  {journal} {Physical review letters}\
  }\textbf {\bibinfo {volume} {82}},\ \bibinfo {pages} {1072} (\bibinfo {year}
  {1999})}\BibitemShut {NoStop}%
\bibitem [{\citenamefont {Trizac}\ and\ \citenamefont
  {Raimbault}(1999)}]{trizac1999long}%
  \BibitemOpen
  \bibfield  {author} {\bibinfo {author} {\bibfnamefont {E.}~\bibnamefont
  {Trizac}}\ and\ \bibinfo {author} {\bibfnamefont {J.-L.}\ \bibnamefont
  {Raimbault}},\ }\bibfield  {title} {\enquote {\bibinfo {title} {Long-range
  electrostatic interactions between like-charged colloids: Steric and
  confinement effects},}\ }\href@noop {} {\bibfield  {journal} {\bibinfo
  {journal} {Physical Review E}\ }\textbf {\bibinfo {volume} {60}},\ \bibinfo
  {pages} {6530} (\bibinfo {year} {1999})}\BibitemShut {NoStop}%
\bibitem [{\citenamefont {Filippov}\ \emph {et~al.}(2006)\citenamefont
  {Filippov}, \citenamefont {Pal}, \citenamefont {Starostin},\ and\
  \citenamefont {Ivanov}}]{filippov2006electrostatic}%
  \BibitemOpen
  \bibfield  {author} {\bibinfo {author} {\bibfnamefont {A.~V.}\ \bibnamefont
  {Filippov}}, \bibinfo {author} {\bibfnamefont {A.}~\bibnamefont {Pal}},
  \bibinfo {author} {\bibfnamefont {A.~N.}\ \bibnamefont {Starostin}}, \ and\
  \bibinfo {author} {\bibfnamefont {A.}~\bibnamefont {Ivanov}},\ }\bibfield
  {title} {\enquote {\bibinfo {title} {Electrostatic interaction between two
  macroparticles in the poisson-boltzmann model},}\ }\href@noop {} {\bibfield
  {journal} {\bibinfo  {journal} {JETP letters}\ }\textbf {\bibinfo {volume}
  {83}},\ \bibinfo {pages} {546--552} (\bibinfo {year} {2006})}\BibitemShut
  {NoStop}%
\bibitem [{\citenamefont {Kolesnikov}, \citenamefont {Mazur},\ and\
  \citenamefont {Budkov}(2022)}]{kolesnikov2022}%
  \BibitemOpen
  \bibfield  {author} {\bibinfo {author} {\bibfnamefont {A.}~\bibnamefont
  {Kolesnikov}}, \bibinfo {author} {\bibfnamefont {D.}~\bibnamefont {Mazur}}, \
  and\ \bibinfo {author} {\bibfnamefont {Y.}~\bibnamefont {Budkov}},\
  }\bibfield  {title} {\enquote {\bibinfo {title} {Electrosorption-induced
  deformation of porous electrode with non-convex pore geometry in electrolyte
  solutions: theoretical study},}\ }\href@noop {} {\bibfield  {journal}
  {\bibinfo  {journal} {to be published}\ } (\bibinfo {year}
  {2022})}\BibitemShut {NoStop}%
\bibitem [{\citenamefont {Derjaguin}, \citenamefont {Churaev},\ and\
  \citenamefont {Muller}(1987)}]{derjaguin1987derjaguin}%
  \BibitemOpen
  \bibfield  {author} {\bibinfo {author} {\bibfnamefont {B.}~\bibnamefont
  {Derjaguin}}, \bibinfo {author} {\bibfnamefont {N.}~\bibnamefont {Churaev}},
  \ and\ \bibinfo {author} {\bibfnamefont {V.}~\bibnamefont {Muller}},\
  }\bibfield  {title} {\enquote {\bibinfo {title} {The
  derjaguin-landau-verwey-overbeek (dlvo) theory of stability of lyophobic
  colloids},}\ }in\ \href@noop {} {\emph {\bibinfo {booktitle} {Surface
  Forces}}}\ (\bibinfo  {publisher} {Springer},\ \bibinfo {year} {1987})\ pp.\
  \bibinfo {pages} {293--310}\BibitemShut {NoStop}%
\bibitem [{\citenamefont {Henderson}(1986)}]{henderson1986compressibility}%
  \BibitemOpen
  \bibfield  {author} {\bibinfo {author} {\bibfnamefont {J.}~\bibnamefont
  {Henderson}},\ }\bibfield  {title} {\enquote {\bibinfo {title}
  {Compressibility route to solvation structure},}\ }\href@noop {} {\bibfield
  {journal} {\bibinfo  {journal} {Molecular Physics}\ }\textbf {\bibinfo
  {volume} {59}},\ \bibinfo {pages} {89--96} (\bibinfo {year}
  {1986})}\BibitemShut {NoStop}%
\end{thebibliography}%
\end{document}